\documentclass[
aps,
prb,
reprint, 
twocolumn,
groupedaddress,
superscriptaddress,
amsfonts,
footinbib,
floatfix,
longbibliography
]{revtex4-2}

\RequirePackage{amsmath,amssymb,bm}
\RequirePackage{graphicx}
\RequirePackage[usenames,dvipsnames]{xcolor}
\usepackage{float}

\usepackage{pdfpages}
\makeatletter\AtBeginDocument{\let\LS@rot\@undefined}\makeatother
\usepackage{pgffor}

\usepackage[T1]{fontenc}

\usepackage{enumitem}
\usepackage{verbatim}

\usepackage{hyperref}
\hypersetup{
    colorlinks=true,
    citecolor= blue,
    linkcolor= blue,
    filecolor=blue,      
    urlcolor= blue,
    pdfauthor = {Chuan Chen},
    pdfcreator = {\LaTeX\ and \flqq hyperref\frqq},
}

\usepackage[capitalise]{cleveref}

\graphicspath{ {./Figs/} }

\newcommand{\Fref}[1]{Fig.~\ref{#1}}
\newcommand{\Frefs}[1]{Figs.~\ref{#1}}
\newcommand{\Eqref}[1]{Eq.~\eqref{#1}}

\newcommand{\bbZ}{\mathbb{Z}}
\newcommand{\bbR}{\mathbb{R}}
\newcommand{\K}{\mathrm{K}}
\newcommand{\M}{\mathrm{M}}
\newcommand{\bk}{\mathbf{k}}
\newcommand{\bq}{\mathbf{q}}
\newcommand{\bQ}{\mathbf{Q}}
\newcommand{\bb}{\mathbf{b}}
\newcommand{\bQIC}{\mathbf{Q}_\mathrm{IC}}
\newcommand{\br}{\mathbf{r}}
\newcommand{\bR}{\mathbf{R}}
\newcommand{\ba}{\mathbf{a}}
\newcommand{\bdelta}{\bm{\delta}}
\newcommand{\scT}{\mathcal{T}}

\makeatletter
\newcommand*{\rom}[1]{\expandafter\@slowromancap\romannumeral #1@}
\makeatother

\begin{document}

\title{Anyon polarons as a window into the competing phases of the Kitaev-Gamma-Gamma-prime model}

\author{Chuan Chen}
\thanks{Corresponding author: chenc@lzu.edu.cn}
\affiliation{School of Physical Science and Technology,
Lanzhou University,
Lanzhou 730000, China}
\affiliation{
Lanzhou Center for Theoretical Physics, 
Key Laboratory of Quantum Theory and Applications of MoE,
Key Laboratory of Theoretical Physics of Gansu Province,
Gansu Provincial Research Center for Basic Disciplines of Quantum Physics,
Lanzhou University,
Lanzhou 730000, China}

\author{Inti Sodemann Villadiego}
\thanks{Corresponding author: sodemann@itp.uni-leipzig.de}
\affiliation{Institut f\"ur Theoretische Physik,
Universit\"at Leipzig,
04103 Leipzig, Germany}

\begin{abstract}
\noindent\textbf{Abstract}

We investigate the dispersions of anyon quasiparticles in the Kitaev honeycomb spin liquid perturbed by $\Gamma$ and $\Gamma'$ couplings
in order to understand phase transitions into competing states through anyon gap-closing instabilities.
We demonstrate how anyon gap closings provide a way to understand phase transitions into a plethora of previously identified competing phases---including
zigzag, stripy, $120^\circ$, and incommensurate spiral phases---and are in agreement with numerical studies not only on the nature of the phases,
but also on the specific critical values of $\Gamma$ and $\Gamma'$ couplings.
Remarkably, when the antiferromagnetic Kitaev model is perturbed by a negative $\Gamma$ interaction, we find that 
the single-vison and fermion gaps remain open while the gap of a magnon-like local boson vanishes.
This points to a plausible candidate coexistence regime where long-range magnetic order emerges while fractionalized excitations inherited 
from the Kitaev spin liquid may persist.
The magnetic order can be either stripy antiferromagnetic or incommensurate spiral, depending on the sign of $\Gamma'$.
\end{abstract}

\date{\today}

\maketitle

\section*{Introduction}
Quantum spin liquids (QSLs) are highly entangled states of matter which feature the emergence of 
fractionalized excitations~\cite{Wen2004,Fradkin_2013,Balents2010-td,Broholm2020-pq}.
Among the various theoretical models proposed to host QSLs~\cite{Rokhsar1988-el,Moessner2001-mz,Kitaev2003-ss,Levin2005,Kitaev2006},
the Kitaev honeycomb model~\cite{Kitaev2006} has emerged as a paradigmatic example due to its exact solvability
and potential realizations~\cite{Jackeli2009-yn} in Kitaev materials such as Na$_2$IrO$_3$, $\alpha$-Li$_2$IrO$_3$, 
and $\alpha$-RuCl$_3$~\cite{Trebst2022-go,Takagi2019-ok,Matsuda2025-hs}.
While the ideal Kitaev model realizes a $\bbZ_2$ QSL with itinerant Majorana fermion (spinon) and static $\bbZ_2$ gauge flux
(vison) excitations, non-Kitaev interactions present in Kitaev materials---such as the Heisenerg exchange ($J$)~\cite{Jackeli2009-yn}, off-diagonal
$\Gamma$ and $\Gamma'$ terms~\cite{Rau2014-ao,Rau2014-ns,Rau2016-oa}, and external magnetic fields---can destabilize the Kitaev spin liquid (KSL) 
and drive the system into other states~\cite{Rau2014-ao,Zhu2018-ve,Gohlke2018-kg,Hickey2019-yn,Gordon2019-iu,Lee2020-zx}. 
In particular, $\alpha$-RuCl$_3$ is believed to be described by a nearest-neighbor $K$-$J$-$\Gamma$-$\Gamma'$ model,
possibly supplemented by third-neighbor Heisenberg interactions ($J_3$) and inter-layer couplings,
although accurately determining these parameters remains a challenge~\cite{Li2021-gy,Maksimov2020-up,Jiefu2025-wi,Moller2025-ly}.
These additional couplings are believed to be responsible for stabilizing the observed zigzag magnetic order at low temperatures, which itself can be suppressed by external magnetic fields 
and replaced by a novel phase---whose nature remains under active debate~\cite{Kasahara2018-or,Yokoi2021-ul,Czajka2021-tm,Czajka2023-da,Bruin2022-wq,Matsuda2025-hs}.
Therefore, understanding the impact of these additional couplings on the ideal KSL remains a major open problem,
the resolution of which could help elucidate more clearly the connections between extended Kitaev models and real materials.

Numerical studies of such extended Kitaev models have uncovered rich phase diagrams
featuring both magnetically ordered phases and other competing novel QSLs (see Refs.~\cite{Trebst2022-go,Rousochatzakis2024-td}
for reviews of the results).
When analyzing extended Kitaev models harboring multiple competing orders,
a powerful starting point is to regard the more symmetric
KSL as the \emph{mother} state, with broken-symmetry phases---such as magnetically ordered states---emerging
as its descendants 
(for earlier studies exploiting this philosophy, see, e.g., Refs.~\cite{Hermele2005-ge,Lee2006-mw}). 

A large class of transitions from the parent QSL to nearby phases can be understood as instabilities where the gap for some of its quasiparticles closes.
In particular, if a QSL hosts a quasiparticle with bosonic self-statistics, its gap closing (i.e., condensation) gives rise to a new phase
in which any other quasiparticle with nontrivial mutual statistics becomes ``confined'',
namely it no longer exists as a finite-energy excitation (see, e.g., Ref.~\cite{Burnell2018-fm}).
For instance, in the toric code $\bbZ_2$ QSL, the condensation of one of its non-local bosons (e.g. the $e$-particle) confines the other anyons, 
resulting in a conventional magnetically ordered state (see, e.g., Ref.~\cite{Pozo2021-gy}).
By contrast, if the condensed quasiparticle is a local boson (i.e., an ordinary magnon-like quasiparticle),
which has trivial statistics with all quasiparticles but carries non-trivial symmetry quantum numbers, 
its gap closing results in a new phase with the same anyon content
of the parent QSL, but with spontaneously broken symmetry dictated by its quantum numbers.
In the case of the ideal Kitaev model, there exists three types of anyonic quasiparticles~\cite{Kitaev2006}:
local bosons, fermions, and a flux particle also known as vison.
Each vison can be viewed as a bare boson dressed by a cloud of fermions since it acts as a $\pi$-flux for the latter.
This ``cloud'' is responsible for transforming the bare bosonic statics of the vison and rendering it, for instance, a non-Abelian anyon when the fermions have
a Bogoliubov Chern band in the presence of a Haldane mass term~\cite{Kitaev2006,Chen2022-tr,Chen2023-vo}.
The gap closing of the single visons is an indication of confinement, which can be viewed as
resulting from the condensation of their bare bosonic core.
Moreover, a pair of visons can form a bound state which can be either a local boson or a fermion.  The fermion vison pairs belong to the same quasi-particle 
super-selection sector as the usual itinerant Majorana fermions of the ideal Kitaev model, therefore, local perturbations can tunnel and mix these quasi-particles, 
and as a result only their combined Chern number remains well defined.
If the fermion gap closes, leading to a change in the fermion Chern number, while the gaps of visons and local bosons remain open, 
the resulting state would generically be another spin liquid belonging to the $16$-fold Kitaev classification~\cite{Kitaev2006}. 
On the other hand, if the lowest-energy quasiparticle is a bosonic vison-pair bound-state and its gap closes under a perturbation without 
the accompanying gap closing of single visons and fermionic particles,
one expects a phase transition from the KSL into a symmetry-broken state that generically retains the same anyon content and topology as the KSL.
In this case the pattern of symmetry breaking can be determined from the quantum number of the condensed bosonic vison-pair.

As will be detailed in latter sections, by computing the dispersions and the quantum numbers of bosonic vison pairs of KSL,
we will be able to determine a variety of magnetic orders observed
in extended Kitaev models, including zigzag, stripy, $120^\circ$ (FM$_{U_6}$ and AFM$_{U_6}$, see definition below)~\cite{Zhang2021-xi,Chen2025-be}.
In most of these cases, we observe that the single visons are also condensed at nearby values of the additional couplings,
indicating that many of the resulting phases are likely trivial magnetically ordered states.
However, in the antiferromagnetic Kitaev model with negative $\Gamma$,
we find a regime where the bosonic visons-pairs condense while single visons remain gapped.
This suggests that the resulting phase is a plausible candidate phase exhibiting coexistence between
long-range magnetic order and fractionalized excitations inherited from the KSL,
with the magnetic order being either stripy or incommensurate (IC) depending on the sign of $\Gamma'$.
Notably, an earlier study by Zhang et al.~\cite{Zhang2021-xi} employed a similar vison-pair approach to analyze the $K$–$J$–$\Gamma$ model
and found results consistent with our findings regarding the $\Gamma$-induced orders, such as the FM$_{U_6}$ ($\Gamma > 0$) and IC ($\Gamma < 0$) orders
in the antiferromagnetic Kitaev model.
However, the single-vison behavior was not considered in this study, leaving open the question of 
whether the resulting phases are trivial magnetically ordered states or QSLs with coexisting symmetry breaking, as we are finding. 
We hope that future numerical studies could identify these interesting phases that combine symmetry breaking with fractionalization,
by investigation of both local magnetic orders and entanglement signatures~\cite{Kitaev2006-ky,Levin2006-zx},
or other indicators such as topological degeneracy in the torus.
 
Moreover, since bosonic vison pairs are created by local operators, they play a central role in understanding typical experimentally measurable
correlation functions of local operators, such as the dynamical spin structure factor $S(\omega,\bQ)$~\cite{Zhang2021-xi}.
Owing to their bosonic character, the bosonic vison pairs can naturally capture the emergence of 
magnon-like low-energy dispersions in $S(\omega,\bQ)$ near the development of magnetic long-range orders.
Finally, we note that the fermionic spinons of the KSL may also undergo gap-closing/opening transitions accompanied by changes 
in their band Chern numbers, which in turn can modify the nature of vison excitations~\cite{Kitaev2006,Chen2025-be,Zhang2022-tc}.
We therefore also analyze in detail the gap closings associated with these fermionic quasiparticles.

Analyzing the dispersion of anyons has provided valuable insights into extended Kitaev models, such as the Kitaev model with 
a Zeeman field~\cite{Chen2023-vo,Chen2025-be,Zhang2022-tc} and the $K$–$J$–$\Gamma$ model~\cite{Zhang2021-xi}.
In this work, we further apply it to the $K$–$\Gamma$–$\Gamma’$ model~\cite{Gordon2019-iu,Chern2020-ti,Lee2020-zx,Sorensen2021-jg} to assess its consistency with numerics and to extract new physical insights.
The model is given by $H_{K \text{-} \Gamma \text{-} \Gamma'} = \sum_{\langle i,j \rangle_{\alpha}} H_{\langle i,j \rangle_{\alpha}}$, with
\begin{align}\label{eq:K-G-G'}
    H_{\langle i,j \rangle_{\alpha}} = & K \, \sigma_i^\alpha \sigma_j^\alpha 
    + \Gamma \,( \sigma_i^\beta \sigma_j^\gamma \nonumber + \sigma_i^\gamma \sigma_j^\beta ) \\
    & + \Gamma' \, ( \sigma_i^\alpha \sigma_j^\beta + \sigma_i^\alpha \sigma_j^\gamma + \sigma_i^\beta \sigma_j^\alpha + \sigma_i^\gamma \sigma_j^\alpha ).
\end{align}
Here $\langle i,j \rangle_{\alpha}$ denotes a bond of type $\alpha=x,y,z$, and $\beta \neq \gamma \neq \alpha$.
Previous numerical studies have revealed several important features of this model.
In the ferromagnetic (FM) Kitaev model [$K = -1$ in \Eqref{eq:K-G-G'}], which is considered pertinent to $\alpha$-RuCl$_3$,
the KSL is highly fragile against the $\Gamma$ interaction. When $\Gamma \gtrsim 0.03$, the KSL has been argued to be
replaced by a novel QSL whose nature remains elusive:
tensor product state and density matrix renormalization group (DMRG) calculations suggest a nematic paramagnetic (NP) phase characterized by anisotropic bond
energies~\cite{Lee2020-zx,Gohlke2020-go}, whereas variational Monte Carlo calculations find a proximate 
KSL phase featuring multiple Majorana cones~\cite{Wang2019-bz,Wang2024-yg}.
Both phases, however, are destabilized in favor of zigzag order upon the inclusion of a small negative $\Gamma' \approx -0.03$~\cite{Lee2020-zx,Wang2024-yg}.
For $\Gamma < 0$, the KSL is replaced by the AFM$_{U_6}$ order (referred to as RS$_{U_6}$ in Ref.~\cite{Sorensen2021-jg}), which maps to a N\'eel order under a six-site
unitary transformation $U_6$~\cite{Chaloupka2015-uh,Sorensen2021-jg,Rousochatzakis2024-td}.
On the other hand, the KSL is more robust against the $\Gamma$ interaction in the antiferromagnetic (AFM) Kitaev model [$K = 1$ in \Eqref{eq:K-G-G'}].
When $\Gamma \gtrsim 0.3$, the KSL is replaced by a FM$_{U_6}$ order which maps to a FM order under the $U_6$
transformation~\cite{Sorensen2021-jg,Rousochatzakis2024-td}. In contrast, for $\Gamma \lesssim -0.5$,
the system could host either stripy or IC spiral order~\cite{Rau2014-ao,Sorensen2021-jg}.
The role of the $\Gamma'$ interaction in this regime is less studied.
The goal of this study is to provide a microscopic understanding of how various phases emerge from the KSL in the $K$-$\Gamma$-$\Gamma'$ model
by starting from the ideal Kitaev limit ($\Gamma = \Gamma' = 0$), computing perturbatively in $\Gamma$ and $\Gamma'$ the dispersions of 
its anyon quasiparticles, and investigating the resulting anyon gap-closing instabilities.


\begin{figure}
\centering
\includegraphics[width=0.45 \textwidth]{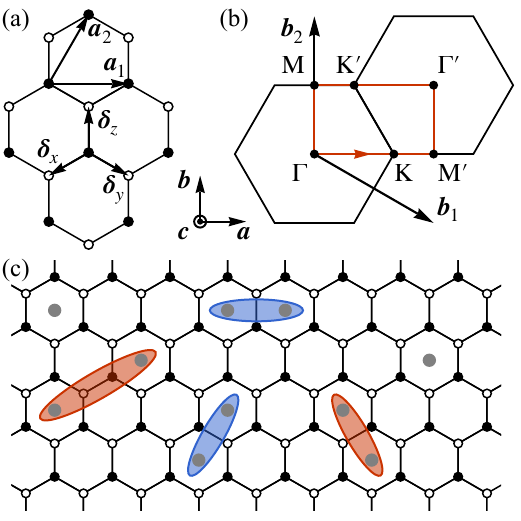}
\caption{
Lattice and momentum-space conventions and quasiparticles of the Kitaev spin liquid.
(a) Honeycomb lattice and its lattice vectors. The $\alpha$-bond of Kitaev model
is along the direction of $\bdelta_{\alpha}$ ($\alpha = x,y,z$).
Filled (open) circles represent the $A$ ($B$) sublattice sites.
The inset illustrates the crystal axis of Kitaev materials.
(b) Brillouin zone (BZ) and the path adopted when calculating the dynamical structure factor 
$S(\omega,\bQ)$ (see \cref{fig:S,fig:S-Kitaev}).
(c) Schematic of visons (gray dots) and vison pairs.
The fermionic vison pairs ($\Tilde{\chi}_{\br,\alpha}$) are represented by blue ellipses,
NN ($d_{\br,\alpha}$) and NNN ($o_{\br,\alpha}$) bosonic vison pairs are represented by red ellipses.
}\label{fig:schematic}
\end{figure}

\section*{Results}
\subsection*{Quasiparticles of the KSL}
Owing to its exact solvability, the Kitaev honeycomb model can be adopted as a well-controlled starting point.
One can therefore start from the KSL and explore, in a perturbative manner, the impact of the $\Gamma$ and $\Gamma’$
interactions on its quasiparticle excitations~\cite{Zhang2021-xi,Zhang2022-tc,Joy2022-ba,Chen2023-vo,Chen2025-be}.
In this work, we focus primarily on the leading (linear) order effects of the $\Gamma$ and $\Gamma'$ interactions.
The KSL contains three types of quasiparticles: single visons, fermions including itinerant Majorana fermions and fermionic vison pairs,
and local bosonic vison pairs.
An important effect of the non-Kitaev interactions like $\Gamma$ and $\Gamma'$ is that they induce fluctuations in the gauge degrees of freedom, 
rendering single visons and vison pairs---both fermionic and bosonic---dynamical, and thereby making the model no longer exactly solvable.
In the following, we analyze the impact of the $\Gamma$ and $\Gamma'$ interactions on each type of quasiparticle in the KSL.

\begin{figure}
\centering
\includegraphics[width=0.5 \textwidth]{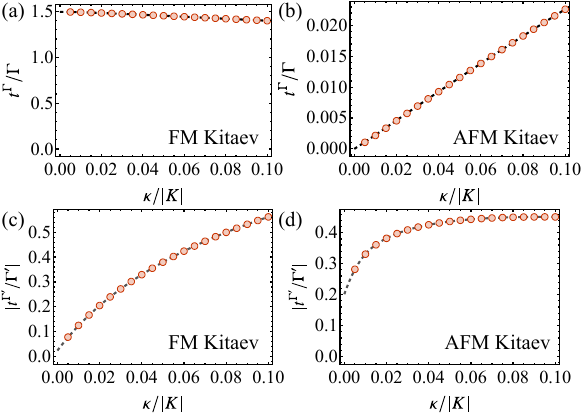}
\caption{
Single-vison hopping induced by the $\Gamma$ and $\Gamma'$ interactions.
The $\Gamma$-induced hopping is finite in the (a) FM Kitaev model, but vanishes in the (b) AFM Kitaev model.
For the $\Gamma'$ interaction, the opposite behavior is found: the induced hopping vanishes in the (c) FM case,
but remains finite in the (d) AFM case.
Note that our calculation yields $t^\Gamma/\Gamma \approx 1.495$, which has the opposite sign to the result
reported in Ref.~\cite{Joy2022-ba}.
A Haldane mass term with coupling $\kappa$ is introduced in the vison hopping calculation, 
and in the $\kappa \rightarrow 0$ limit, the vison separation is kept greater than $\xi_c = v_c/\Delta_c$.
}\label{fig:t-sv}
\end{figure}

The single visons move on an effective triangular lattice formed by the hexagons of the honeycomb lattice
(see \Fref{fig:schematic}(c))~\cite{Joy2022-ba,Chen2023-vo}.
The $\Gamma$/$\Gamma'$ interaction enables a single vison to hop between next-nearest-neighbor (NNN)/nearest-neighbor (NN)
hexagons.
To linear order in $\Gamma$/$\Gamma'$, due to the existence of a vison gap $\Delta_v$,
the effective Hamiltonian of a single vison only contains hoppings within the degenerate states
associated with its position.
Therefore, the effective Hamiltonian for single visons has the form:
\begin{align}\label{eq:H_sv}
    H_\mathrm{v} = \sum_{\langle i,j \rangle} t_{i,j}^{\Gamma'} |v_i\rangle \langle v_j|
    + \sum_{\langle \! \langle i,j \rangle \! \rangle} t_{i,j}^{\Gamma} |v_i \rangle \langle v_j|
    + \sum_i \Delta_v |v_i \rangle \langle v_i|.
\end{align}
Here $|v_i \rangle$ is the state with a vison at plaquette $i$.
$t_{i,j}^{\Gamma/\Gamma'}$ is the $\Gamma$/$\Gamma'$-induced hopping amplitude,
defined as the matrix element of the $\Gamma/\Gamma'$ interaction
between states where a vison locates at sites $i$ and $j$ (see its expression in Supplementary Information).
$\Delta_v \approx 0.15 |K|$ is the single-vison creation energy~\cite{Kitaev2006}.

\begin{figure}
\centering
\includegraphics[width=0.46 \textwidth]{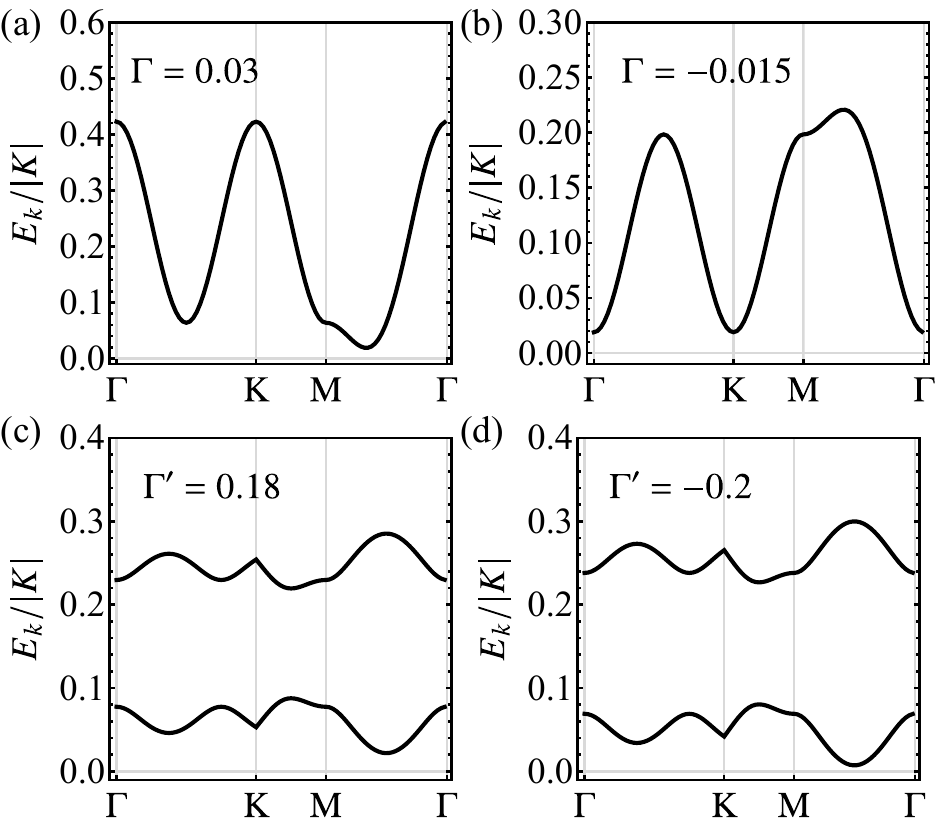}
\caption{
Single-vison bands for the FM and AFM Kitaev models.
(a)--(b) Single-vison bands for the FM Kitaev model with $\Gamma = 0.03$ and $\Gamma = -0.015$ respectively.
(c)--(d) Single-vison bands for the AFM Kitaev model with $\Gamma' = 0.18$ and $\Gamma' = -0.2$ respectively.
The presence of two bands in this case arises from the fractionalized translational symmetry,
where each vison effectively experiences a background $\pi$ flux per honeycomb unit cell.
Here we plot the bands using the same momenta as in the FM case to facilitate comparison.
The high-symmetry line in the honeycomb lattice BZ is illustrated in \Fref{fig:schematic}(b).
}\label{fig:sv-band}
\end{figure}

To define the single-vison hopping in the physical Hilbert space, where visons are created in pairs, 
we place an auxiliary second vison far away from the local perturbation. 
Using the computational methods developed in Refs.~\cite{Chen2022-tr,Chen2023-vo,Chen2025-be,Joy2022-ba},
we find that in the FM Kitaev model, $t^\Gamma/\Gamma$ remains finite whereas $t^{\Gamma'}/\Gamma'=0$.
By contrast, in the AFM Kitaev model, $t^\Gamma/\Gamma=0$ whereas $t^{\Gamma'}/\Gamma'$ remains finite,
in agreement with Ref.~\cite{Joy2022-ba} (see \Fref{fig:t-sv}).
The vanishing leading-order $t^\Gamma/\Gamma$ ($t^{\Gamma'}/\Gamma'$) in the AFM (FM) Kitaev model
can be understood as arising from destructive interference among multiple symmetry-related single-vison hopping channels
(see Supplementary Information for details).
Moreover, our computation reveals that the $t^{\Gamma'}_{i,j}$ values 
give rise to a non-trivial projective implementation of translational symmetry,
corresponding to a $\pi$-flux per honeycomb unit cell experienced by visons in the AFM Kitaev model,
thereby effectively doubling their unit cell (see vison bands in \Fref{fig:sv-band}).
As we have shown previously~\cite{Chen2022-tr,Chen2023-vo,Chen2025-be}, this pattern of translational symmetry 
fractionalization is an intrinsic property of the AFM-KSL, independent of the perturbation that induces vison mobility, 
making it a sharply distinct symmetry enriched topologically ordered state compared to the FM-KSL,
whose visons have a trivial implementation of lattice translations.
The significant (vanishing) value of $t^\Gamma/\Gamma$ in the FM (AFM) Kitaev model implies that the single-vison gap closes at relatively
small (large) $\Gamma$, consistent with the weaker (stronger) stability of KSL against $\Gamma$ interaction in the FM (AFM) case~\cite{Rau2014-ao,Zhang2021-xi}.
We have also examined the second-order single-vison hopping $t^{\Gamma,(2)}$ induced by $\Gamma$, and found a similar
qualitative behavior: $t^{\Gamma,(2)}/\Gamma^2 \neq 0$ ($= 0$) for FM (AFM) Kitaev model 
(see Supplementary Information for details). 
This supports the idea that, in the AFM Kitaev model, single visons remain gapped---that is, matter fermions remain deconfined---until much larger $\Gamma$ values,
opening the door to a novel $\Gamma$-induced QSL phase exhibiting spontaneous symmetry breaking (see discussions in later sections).
Although the linear-order single-vison hopping induced by the $\Gamma'$ interaction vanishes in the FM Kitaev model,
higher-order contributions are expected to render it finite.

Crucially, unlike the case of a Zeeman field---which induces a Haldane mass gap on the matter fermions~\cite{Kitaev2003-ss},
and renders the vison an exponentially localized quasiparticle ($\sigma$ anyon of the Ising topological order)---the $\Gamma$ and $\Gamma'$ interactions
do not immediately gap out the fermions. 
In the present context, this implies that visons behaves as vortices which are dressed by power-law decaying
modification of the surrounding matter fermion fluid. 
Therefore, we need to exercise caution to correctly extract their hopping amplitude numerically
since the states in a torus always contain at least two visons.
To do so, we have introduced an artificial Haldane mass ($\kappa$) term to gap out the matter fermions,
allowing us to define visons as well-separated individual excitations.
We then extract the thermodynamic-limit values of $t^{\Gamma}/\Gamma$ and $t^{\Gamma'}/\Gamma'$ by taking the limit $\kappa \rightarrow 0$,
while keeping the vison–vison separation remains greater than their effective localization length $\xi_c = v_c/\Delta_c$, where $v_c = \sqrt{3}|K|$ is the $c$-fermion Fermi velocity
and $\Delta_c = 6\sqrt{3}\kappa$ is the gap induced by the mass term.

In contrast to an isolated vison, which acts as a $\pi$-flux defect for the matter fermions, 
a nearby vison pair formed by two NN or NNN visons constitutes a local perturbation,
since the total flux seen by distant matter fermions is $0 \ \mathrm{mod}\ 2\pi$.
When the matter fermions are in their ground state in a given flux sector, two nearby visons typically have a lower energy than two well-separated visons
(see the two-vison excitation energy as a function of their separation in Ref.~\cite{Chen2025-be}),
and induce a more rapidly decaying power-law modification of the surrounding matter-fermion fluid~\cite{Zhang2021-xi}.
This localized dressing cloud provides the physical basis for treating low-energy fermionic vison pairs,
which correspond to the matter-fermion ground state in the two-vison sector (see \cref{eq:fvp-def}), as coherent quasiparticles.

Bosonic vison-pair states represent a distinct quasiparticle channel, since they additionally involve a matter-fermion Bogoliubov quasiparticle 
in the background of two nearby visons (see \cref{eq:d_r-def}) and are not necessarily the lowest-energy bound states in that sector.
In this work, we treat them as quasiparticle channels within our effective nearby-vison-pair description.
When such a bosonic mode becomes low in energy, it provides a natural instability channel toward a symmetry-broken phase.
At the same time, compared with the low-energy fermionic vison pairs, these bosonic channels may be more susceptible to higher-order effects, 
such as hybridization with separated-vison states or vison-pair dissociation.
Below, we analyze the properties of these vison-pair channels in the presence of the $\Gamma$ and $\Gamma'$ interactions.
%

\begin{figure}
\centering
\includegraphics[width=0.5 \textwidth]{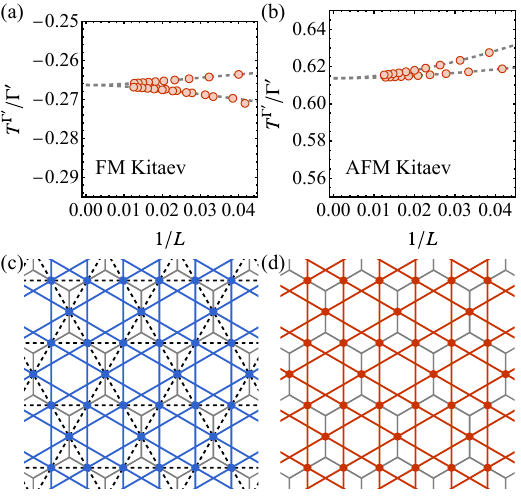}
\caption{
$\Gamma'$-induced $\Tilde{\chi}$-fermion hopping and the effective tight-binding lattice for vison pairs.
(a)--(b) Finite-size scaling of $\Gamma'$-induced $\Tilde{\chi}$-fermion hopping for the FM and AFM Kitaev models.
For results on $\Gamma$-induced hopping $T^{\Gamma}$, see Ref.~\cite{Zhang2021-xi}.
(c) Illustration of the hopping processes of NN vison pairs (both fermionic and bosonic).
Vison pairs (blue dots) reside on the bonds of the original honeycomb lattice (gray) and they form a Kagome lattice.
$\Gamma$-induced NNN hopping is shown as blue lines, and $\Gamma'$-induced NN hopping as black dashed lines.
(d) Hopping processes of NNN bosonic vison pairs (red dots), with $\Gamma$-induced hopping shown as red lines.
}\label{fig:t-f}
\end{figure}

\begin{figure*}
\centering
\includegraphics[width=1 \textwidth]{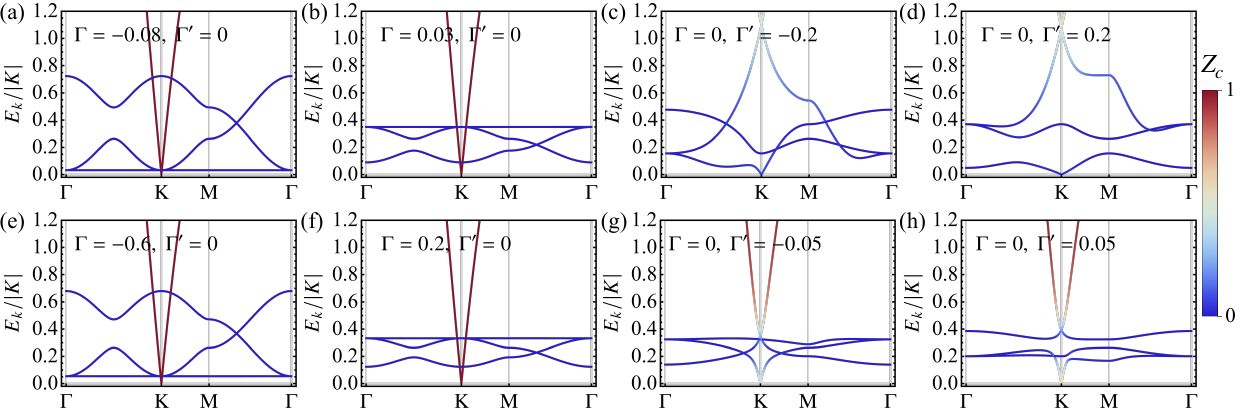}
\caption{
Fermionic quasiparticle bands induced by the $\Gamma$ and $\Gamma'$ interactions.
(a)--(d) Fermion bands for the FM Kitaev model at selected points.
(e)--(h) Fermion bands for the AFM Kitaev model at selected points.
In the absence of the $\Gamma'$ interaction, the $\Tilde{\chi}$ vison pairs and the $c$-Majorana fermions remain decoupled.
As a result, the $c$-fermion bands remain unchanged in panels (a), (b), (e), and (f),
while the flat $\Tilde{\chi}$ bands originate from the effective Kagome geometry of the corresponding tight-binding model.
When $\Gamma'$ is finite, the structure of $p_{\bR,\alpha}$ ensures that the spectrum remains gapless at the $\K$ point.
The color indicates the $c$-fermion weight $Z_c$ of each eigenmode.
}\label{fig:f-band}
\end{figure*}

A fermionic vison pair consists of two NN visons (see the illustration in \Fref{fig:schematic}(c)) accompanied by a modified matter fermion
background~\cite{Knolle2014-wl,Knolle2015-lx,Zhang2021-xi,Zhang2022-tc,Chen2025-be}.
Utilizing the Majorana representation of local spins ($\sigma_{\br}^\alpha \leftrightarrow i b_{\br}^\alpha c_{\br}$)~\cite{Kitaev2006},
a fermionic vison pair on bond $(\br,\alpha)$ with $\br \in A$ is defined as:
\begin{align}\label{eq:fvp-def}
    \Tilde{\chi}_{\br,\alpha}^{\dagger}|\Omega\rangle = \chi_{\br,\alpha}^\dagger | 0; \Psi_c(\br,\alpha) \rangle.
\end{align}
Here $|\Omega\rangle = |0;\Psi_c(0) \rangle$ denotes the ground state of the Kitaev Hamiltonian without any flux.
$\chi_{\br,\alpha}^\dagger = b_{\br}^\alpha - i b_{\br+\bdelta_{\alpha}}^\alpha$ is
the creation operator of a $\bbZ_2$ gauge fermion defined on bond $(\br,\alpha)$.
The state $|0, \Psi_c(\br,\alpha) \rangle$ is the ground state of a matter $c$-fermion BdG Hamiltonian $H_c(\br,\alpha)$ 
incorporating a pair of $\pi$ fluxes separated by the bond $(\br,\alpha)$.
Importantly, $|0, \Psi_c(\br,\alpha) \rangle$ has the same (even) total fermion parity as $| \Omega \rangle$, 
so $\Tilde{\chi}_{\br,\alpha}^\dagger|\Omega \rangle$ carries the opposite (odd) fermion parity
and is therefore referred to as a fermionic vison pair~\cite{Zschocke2015-zr,Knolle2015-lx,Zhang2021-xi,Chen2023-vo}.
For notational simplicity, the state defined in \cref{eq:fvp-def} should be understood as a shorthand expression in the enlarged Majorana Hilbert space;
the corresponding physical state is obtained by applying the appropriate projection operator.
For the fermionic vison-pair state, this involves a modified projection in which one site far from the bond $(\br,\alpha)$
carries the opposite local gauge constraint, which is equivalent to introducing a distant gauge fermion.
The explicit projected expression and its gauge-invariant interpretation are given in the Supplementary Information.
%

The $\Gamma$ interaction induces hopping of $\Tilde{\chi}$-fermions~\cite{Zhang2021-xi}, while the $\Gamma'$ interaction 
not only generates additional $\Tilde{\chi}$-hopping but also leads to hybridization between the $\Tilde{\chi}$ and $c$ fermions (see \Fref{fig:t-f}(c)).
Specifically, the $\Gamma'$ interaction gives rise to tunneling as well as pair creation and annihilation of $\Tilde{\chi}$ and $c$ fermions
(see Supplementary Information).
The effective Hamiltonian of the fermion quasiparticles is therefore:
\begin{align}\label{eq:H_f}
    H_\mathrm{f} = H_c(0) + \sum_{\br \in A,\alpha} \Delta_{\Tilde{\chi}} \Tilde{\chi}_{\br,\alpha}^{\dagger} \Tilde{\chi}_{\br,\alpha} 
    + H^\Gamma_\mathrm{f} + H^{\Gamma'}_\mathrm{f}.
\end{align}
Here $H_c(0)$ is the $c$-fermion BdG Hamiltonian in the zero-flux sector, $\Delta_{\Tilde{\chi}} \approx 0.26|K|$ is the excitation gap
of a fermion vison pair in Kitaev model~\cite{Chen2025-be,Zhang2021-xi}. The part induced by $\Gamma$ and $\Gamma'$ interactions reads:
\begin{align}\label{eq:H^G_f}
    H^\Gamma_\mathrm{f} & = \sum_{\br \in A} \left[ T^{\Gamma,zy}_{-\ba_2} \Tilde{\chi}_{\br,z}^\dagger \Tilde{\chi}_{\br+\ba_2,y}
    + T^{\Gamma,xz}_{\ba_1} \Tilde{\chi}_{\br,x}^\dagger \Tilde{\chi}_{\br-\ba_1,z} \right. \nonumber \\
    & + T^{\Gamma,yx}_{-\ba_1+\ba_2} \Tilde{\chi}_{\br,y}^\dagger \Tilde{\chi}_{\br+\ba_1-\ba_2,x} + 
    T^{\Gamma,zy}_{\ba_1} \Tilde{\chi}_{\br,z}^\dagger \Tilde{\chi}_{\br-\ba_1,y} \nonumber \\
    & \left. + T^{\Gamma,xz}_{-\ba_1+\ba_2} \Tilde{\chi}_{\br,x}^\dagger \Tilde{\chi}_{\br+\ba_1-\ba_2,z} +
    T^{\Gamma,yx}_{-\ba_2} \Tilde{\chi}_{\br,y}^\dagger \Tilde{\chi}_{\br+\ba_2,x} + \mathrm{H.c.} \right],
\end{align}
\begin{align}\label{H^G'_f}
    H^{\Gamma'}_{\mathrm{f}} & = \sum_{\br \in A} \left[ T^{\Gamma',yz}_{-\ba_1+\ba_2} \Tilde{\chi}_{\br,y}^\dagger \Tilde{\chi}_{\br+\ba_1-\ba_2,z}
    + T^{\Gamma',xy}_{\ba_1} \Tilde{\chi}_{\br,x}^\dagger \Tilde{\chi}_{\br-\ba_1,y} \right. \nonumber \\
    & + T^{\Gamma',zx}_{-\ba_2} \Tilde{\chi}_{\br,z}^\dagger \Tilde{\chi}_{\br+\ba_2,x} + 
    T^{\Gamma',yz}_{0} \Tilde{\chi}_{\br,y}^\dagger \Tilde{\chi}_{\br,z} \nonumber \\
    & \left. + T^{\Gamma',xy}_{0} \Tilde{\chi}_{\br,x}^\dagger \Tilde{\chi}_{\br,y} + T^{\Gamma',zx}_{0} \Tilde{\chi}_{\br,z}^\dagger \Tilde{\chi}_{\br,x}
    + \mathrm{H.c.} \right] \nonumber \\
    & + \sum_{\br \in A, \alpha} \sum_{\bR} p_{\bR,\alpha} ( c_{\br+\bR} - i c_{\br+\bdelta_\alpha-\bR} ) \Tilde{\chi}_{\br,\alpha} + \mathrm{H.c.}
\end{align}
The fermion hopping parameter $T^{\Gamma/\Gamma',\alpha \beta}_{-\bR}$ ($\bR$ is a lattice vector) is defined as the matrix element of 
the $\Gamma/\Gamma'$ interaction between states $\langle \Omega | \Tilde{\chi}_{\br,\alpha}$ and $\Tilde{\chi}_{\br+\bR,\beta}^\dagger | \Omega \rangle$. 
E.g.,
\begin{align}
    T^{\Gamma,zy}_{-\ba_2} & = \langle \Omega| \Tilde{\chi}_{\br,z} H^{\Gamma}_{\br+\ba_2,x} \Tilde{\chi}_{\br+\ba_2,y}^\dagger
    |\Omega \rangle \nonumber \\
    & = -\Gamma \langle 0; \Psi_c(\br,z)| \nonumber \\
    & \left( 1 + i\, c_{\br+\ba_2} c_{\br+\bdelta_z} \right) |0;\Psi_c(\br+\ba_2,y) \rangle,
\end{align}
\begin{align}\label{eq:Gmp-chi-hop}
    T^{\Gamma',yz}_0 & = \langle \Omega| \Tilde{\chi}_{\br,y}
    \left[ H^{\Gamma'}_{\br,z} + H^{\Gamma'}_{\br,y} \right. \nonumber \\
    & \left. + H^{\Gamma'}_{\br-\ba_1,y} + H^{\Gamma'}_{\br-\ba_2,z} \right]
    \Tilde{\chi}_{\br,z}^\dagger | \Omega \rangle
    \nonumber \\ 
    & = \Gamma' \langle 0;\Psi_c(\br,y)| c_{\br} \, g |0;\Psi_c(\br,z) \rangle, \nonumber \\
    g & = c_{\br-\ba_2} - c_{\br-\ba_1} - i c_{\br+\bdelta_z} - i c_{\br+\bdelta_y}.
\end{align}
Here $H^{\Gamma/\Gamma'}_{\br,\alpha}$ stands for the $\Gamma/\Gamma'$ interaction on bond $(\br,\alpha)$.
Note for each $H^{\Gamma'}_{\br,\alpha}$ in \Eqref{eq:Gmp-chi-hop}, there is only one two-spin
term contributes to the hopping process.
Due to rotational symmetry, the $\Gamma$-induced $\Tilde{\chi}$-fermion hopping amplitudes satisfy:
\begin{align}\label{eq:T^G_f}
    T^{\Gamma,zy}_{-\ba_2} & = T^{\Gamma,xz}_{\ba_1} =  T^{\Gamma,yx}_{-\ba_1+\ba_2}
    = T^{\Gamma,zy}_{\ba_1} = T^{\Gamma,xz}_{-\ba_1+\ba_2} = T^{\Gamma,yx}_{-\ba_2} \nonumber \\
    & = T^\Gamma \in \bbR,
\end{align}
and $T^\Gamma/\Gamma \approx -1.44 \ (-0.17)$ in the FM (AFM) Kitaev model (see Ref.~\cite{Zhang2021-xi} for plots of $T^\Gamma/\Gamma$ at different system sizes).
Similarly, these same symmetries constraint the $\Gamma'$ terms as follows:
\begin{align}\label{T^G'_f}
    T^{\Gamma',yz}_{-\ba_1+\ba_2} & = T^{\Gamma',xy}_{\ba_1} = T^{\Gamma',zx}_{-\ba_2} 
    = T^{\Gamma',yz}_{0} = T^{\Gamma',xy}_{0} = T^{\Gamma',zx}_{0} \nonumber \\
    & = T^{\Gamma'} \in \bbR,
\end{align}
with $T^{\Gamma'}/\Gamma' \approx -0.27 \ (0.61)$ in the FM (AFM) Kitaev model (see \Frefs{fig:t-f}(a)--(b)).
Note that the $\Gamma$-induced fermion hopping in the FM Kitaev model is an order of magnitude stronger
than in the AFM Kitev model. This suggests that the fermions are more likely to undergo gap closing at much smaller $\Gamma$
values in the FM Kitaev model, consistent with the greater fragility of the KSL under this perturbation.
By contrast, the $\Gamma'$-induced fermion hopping is approximately twice as strong in the AFM model compared to the FM one.

While $T^{\Gamma}$ and $T^{\Gamma'}$ exhibit strong dependence on the sign of $K$,
the $\Tilde{\chi}$-$c$ coupling $p_{\bR,\alpha}$ is found to be independent of it.
Moreover, a $\Tilde{\chi}_{\br,\alpha}$ fermion couples only to nearby $c$ fermions,
with $\max(|p_{\bR,\alpha}/\Gamma'|) \approx 0.88$ (see Supplementary Information).

\begin{figure*}
\centering
\includegraphics[width=1 \textwidth]{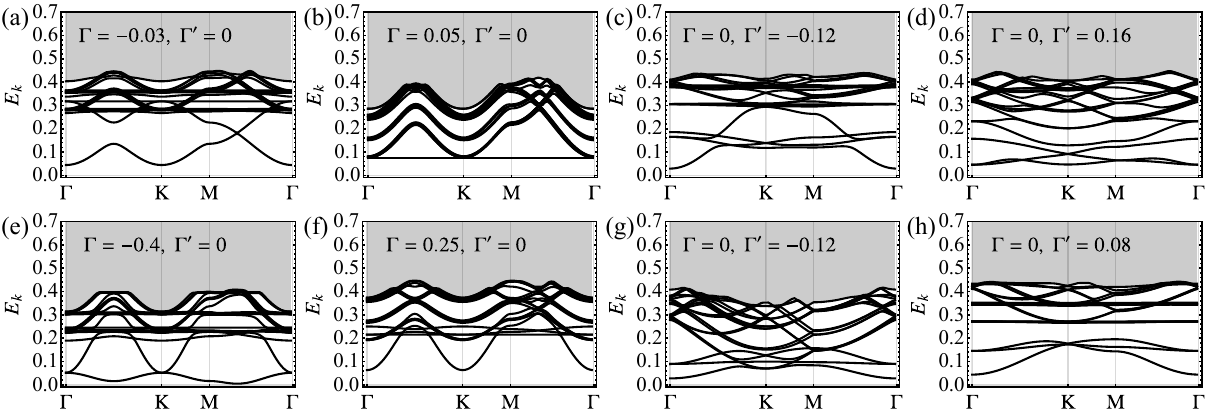}
\caption{
Bosonic quasiparticle bands induced by the $\Gamma$ and $\Gamma'$ interactions.
(a)--(d) Boson bands of the FM Kitaev model at representative points.
(e)--(h) Boson bands of the AFM Kitaev model at representative points.
Only the lowest $20$ bands are shown, whereas other higher-energy bands are indicated by the gray shading.
When $\Gamma' = 0$, the NN and NNN bosonic vison pairs are decoupled. 
In the FM Kitaev model with $\Gamma' = 0$, the lowest band originates from the NNN vison pairs, so the soft boson mode does not induce local magnetic order.
In contrast, in the AFM Kitaev model, the lowest band arises from the NN vison pairs, and the condensation of soft modes 
gives rise to magnetic long-range order. Results here are based on calculations performed on a $60 \times 60$ lattice.
}\label{fig:b-band}
\end{figure*}

A NN bosonic vison pair at bond $(\br,\alpha)$ consists of a fermion vison pair and a matter fermion excitation~\cite{Zhang2021-xi,Chen2023-vo}:
\begin{align}\label{eq:d_r-def}
    d_{\br,\alpha,l}^\dagger | \Omega \rangle = \chi_{\br,\alpha}^\dagger \alpha_{(\br,\alpha),l}^\dagger | 0; \Psi_c(\br,\alpha) \rangle.
\end{align}
Here $\alpha_{(\br,\alpha),l}^\dagger$ is the creation operator for the $l$-th ($l = 1,\dots, N$, with $N$ being the number
of unit cells in the system) Bogoliubov quasiparticle of the matter fermion
Hamiltonian $H_c(\br,\alpha)$, and $d_{\br,\alpha,l}^\dagger$ denotes the creation operator for the $l$-th NN bosonic vison pair
at bond $(\br,\alpha)$.
An important feature of the vison pair $d_{\br,\alpha,l}^\dagger$ is that it can be created from, or annihilated into,
the KSL vacuum $|\Omega \rangle$ by local operators, such as Pauli operators $\sigma_{\br}^\alpha$
and $\sigma_{\br+\bdelta_\alpha}^\alpha$ or their products.
These quasiparticles can be therefore viewed as generalized magnons whose 
condensation can induce spontaneous symmetry breaking and magnetic orders,
as formulated in the Methods~\cite{Chen2023-vo,Zhang2021-xi}. 
They also play an important role in the behavior of correlation functions of local operators,
such as the dynamical spin structure factor $S(\omega,\bQ)$
(see its expression in the Methods)~\cite{Knolle2014-wl,Knolle2015-lx,Zhang2021-xi}.
As $S(\omega,\bQ)$ in the ideal Kitaev model is quantitatively well reproduced by the bosonic vison-pair 
approximation (see plots in \Fref{fig:S-Kitaev})~\cite{Zhang2021-xi,Knolle2015-lx}, 
extending this framework to extended Kitaev models
can provide valuable insights into emergent magnetic orders driven by non-Kitaev interactions.

In addition to the NN bosons, we also consider the NNN bosonic vison pairs, as they have slightly lower energy than the NN vison pairs in the ideal
Kitaev model (with excitation energy $\Delta_o \approx 0.24|K|$)~\cite{Kitaev2006}.
A NNN vison pair with two visons being connected by the bond $(\br,\alpha)$ is defined as (see illustration in \Fref{fig:schematic}(c)):
\begin{align}\label{eq:o_r-def}
    o_{\br,\alpha}^\dagger |\Omega \rangle = \chi_{\br+\bdelta_\alpha-\bdelta_\gamma,\gamma}^\dagger \chi_{\br,\beta}^\dagger  
    |0;\Psi_c(\br+\bdelta_\alpha-\bdelta_\gamma,\gamma; \br,\beta)\rangle.
\end{align}
Here $o_{\br,\alpha}^\dagger$ is the creation operator for the NNN boson vison pair at bond $(\br,\alpha)$.
In \Eqref{eq:o_r-def}, the indices $\beta$ and $\gamma$ are fixed by $\alpha$ through cyclic ordering.
That is, $(\alpha, \beta, \gamma)$ always form an ordered cyclic permutation of $(x,y,z)$;
e.g., if $\alpha = z$, then $\beta = x$ and $\gamma = y$.
The state $|0;\Psi_c(\br+\bdelta_\alpha-\bdelta_\gamma,\gamma; \br,\beta)\rangle$ is the ground state of matter fermion
Hamiltonian $H_c(\br+\bdelta_\alpha-\bdelta_\gamma,\gamma; \br,\beta)$, which hosts two $\pi$ fluxes connected by $(\br,\alpha)$.
$o_{\br,\alpha}^\dagger |\Omega\rangle$ is bosonic since it shares the same fermion parity as $|\Omega \rangle$.

The $\Gamma$ interaction induces hopping of both NN and NNN bosons, whereas the $\Gamma'$ interaction generates hopping
of NN bosons and tunneling between NN and NNN bosons, as illustrated in \Frefs{fig:t-f}(c)--(d).
The boson effective Hamiltonian reads: $H_\mathrm{b} = H_\mathrm{b}^d + H_{\mathrm{b}}^o + H_{\mathrm{b}}^{d o}$, with
\begin{align}\label{H_b^d}
    H_\mathrm{b}^d & = \sum_{\br \in A, \alpha} d_{\br,\alpha}^\dagger \Delta_{d} d_{\br,\alpha} \nonumber \\
    & + \sum_{\br \in A} \left[ M^{\Gamma,zy}_{-\ba_2} d_{\br,z}^\dagger d_{\br+\ba_2,y}
    + M^{\Gamma,xz}_{\ba_1} d_{\br,x}^\dagger d_{\br-\ba_1,z} \right. \nonumber \\
    & + M^{\Gamma,yx}_{-\ba_1+\ba_2} d_{\br,y}^\dagger d_{\br+\ba_1-\ba_2,x} + 
    M^{\Gamma,zy}_{\ba_1} d_{\br,z}^\dagger d_{\br-\ba_1,y} \nonumber \\
    & + M^{\Gamma,xz}_{-\ba_1+\ba_2} d_{\br,x}^\dagger d_{\br+\ba_1-\ba_2,z} +
    M^{\Gamma,yx}_{-\ba_2} d_{\br,y}^\dagger d_{\br+\ba_2,x} \nonumber \\
    & + M^{\Gamma',yz}_{-\ba_1+\ba_2} d_{\br,y}^\dagger d_{\br+\ba_1-\ba_2,z}
    + M^{\Gamma',xy}_{\ba_1} d_{\br,x}^\dagger d_{\br-\ba_1,y} \nonumber \\
    & + M^{\Gamma',zx}_{-\ba_2} d_{\br,z}^\dagger d_{\br+\ba_2,x} + 
    M^{\Gamma',yz}_{0} d_{\br,y}^\dagger d_{\br,z} \nonumber \\
    & \left. + M^{\Gamma',xy}_{0} d_{\br,x}^\dagger d_{\br,y} + M^{\Gamma',zx}_{0} d_{\br,z}^\dagger d_{\br,x} + \mathrm{H.c.} \right],
\end{align}
\begin{align}\label{H_b^o}
    H_\mathrm{b}^o & = \sum_{\br \in A, \alpha} \Delta_o o_{\br,\alpha}^\dagger o_{\br, \alpha} \nonumber \\
    & + \sum_{\br \in A} \left[ t^{o,zx}_{-\ba_1} o_{\br,z}^\dagger o_{\br+\ba_1,x} + t^{o,yz}_{-\ba_1} o_{\br,y}^\dagger o_{\br+\ba_1,z} \right. \nonumber \\
    & + t^{o,xy}_{\ba_1-\ba_2} o_{\br,x}^\dagger o_{\br-\ba_1+\ba_2,y} + t^{o,zx}_{\ba_1-\ba_2} o_{\br,z}^\dagger o_{\br-\ba_1+\ba_2,x} \nonumber \\
    & + \left. t^{o,yz}_{\ba_2} o_{\br,y}^\dagger o_{\br-\ba_2,z} + t^{o,xy}_{\ba_2} o_{\br,x}^\dagger o_{\br-\ba_2,y} + \mathrm{H.c.} \right],
\end{align}
\begin{align}\label{H_b^{do}}
    H_\mathrm{b}^{d o} & = \sum_{\br \in A} o_{\br,x}^\dagger ( \scT^{x}_1 d_{\br,y} + \scT^{x}_2 d_{\br,z} + \scT^{x}_3 d_{\br-\ba_1,y} \nonumber \\
    & + \scT^{x}_4 d_{\br-\ba_2,z} ) + o_{\br,y}^\dagger ( \scT^{y}_1 d_{\br,z} + \scT^{y}_2 d_{\br,x} \nonumber \\
    & +  \scT^{y}_3 d_{\br+\ba_1-\ba_2,z} + \scT^{y}_4 d_{\br+\ba_1,x} ) + o_{\br,z}^\dagger ( \scT^{z}_1 d_{\br,x} \nonumber \\
    & + \scT^{z}_2 d_{\br,y} + \scT^z_{3} d_{\br+\ba_2,x} + \scT^z_4 d_{\br-\ba_1+\ba_2,y} ) + \mathrm{H.c.}
\end{align}
Here we have introduced the compact notation 
\begin{equation}\label{eq:d-compact}
    d_{\br,\alpha}^\dagger = (d_{\br,\alpha,1}^\dagger, \dots, d_{\br,\alpha,N}^\dagger).
\end{equation}
The diagonal matrix $\Delta_{d}$ contains the excitation energies $\Delta_{d,l}$ of $d_{r,\alpha,l}$ bosons.
The $N \times N$ hopping matrix $M^{\Gamma/\Gamma'}$ is a generalization of the fermion hopping parameter $T^{\Gamma,\Gamma'}$,
where $[ M^{\Gamma/\Gamma',\alpha \beta}_{-\bR} ]_{m,n}$ is the matrix element of $\Gamma/\Gamma'$ interaction between states
$\langle \Omega| d_{\br,\alpha,m}$ and $d_{\br+\bR,\beta,n}^\dagger|\Omega \rangle$. E.g.,
\begin{align}\label{eq:M^G}
    [M^{\Gamma,zy}_{-\ba_2}]_{m,n} & = \langle \Omega| d_{\br,z,m} H^{\Gamma}_{\br+\ba_2,x}
    d_{\br+\ba_2,y,n}^\dagger |\Omega \rangle \nonumber \\
    & = -\Gamma  \langle 0; \Psi_c(\br,z)| \alpha_{(\br,z),m} \left[ 1 + i c_{\br+\ba_2} c_{\br+\bdelta_z} \right] 
    \nonumber \\
    & \  \alpha_{(\br+\ba_2,y),n}^\dagger |0; \Psi_c(\br+\ba_2,y) \rangle,
\end{align}
\begin{align}\label{eq:M^G'}
    [M^{\Gamma',yz}_{0}]_{m,n} & = \langle \Omega| d_{\br,y,m} \left[ H^{\Gamma'}_{\br,z} + H^{\Gamma'}_{\br,y}
    \right. \nonumber \\
    & \left. + H^{\Gamma'}_{\br-\ba_2,z} + H^{\Gamma'}_{\br-\ba_1,y} \right] d_{\br,z,n}^\dagger |\Omega \rangle \nonumber \\
    & = \Gamma' \langle 0;\Psi_c(\br,y)| \alpha_{(\br,y),m} c_{\br} \left[ c_{\br-\ba_2} - c_{\br-\ba_1} \right. \nonumber\\
    & \left. - i c_{\br+\bdelta_z} - i c_{\br+\bdelta_y} \right] \alpha_{(\br,z),m}^\dagger |0;\Psi_c(\br,z) \rangle.
\end{align}
The $1 \times N$ tunneling vector $\scT^{\alpha}$ is defined as the matrix element of $\Gamma'$ interaction between NN and NNN boson states. E.g.,
\begin{align}\label{eq:T^x-eg}
    [ \scT^x_1 ]_n & = \langle \Omega| o_{\br,x} \left[ H^{\Gamma'}_{\br,x} + H^{\Gamma'}_{\br-\ba_1,y} \right. \nonumber \\
    & \left. + H^{\Gamma'}_{\br-\ba_2,x} + H^{\Gamma'}_{\br-\ba_2,y} \right] d_{\br,y,n}^\dagger | \Omega \rangle \nonumber \\
    & = -\Gamma' \langle 0;\Psi_c(\br-\ba_2,z; \br,y)| g \alpha_{(\br,y),n}^\dagger |0; \Psi_c(\br,y) \rangle, \nonumber \\
    g & = c_{\br} - c_{\br-\ba_1} - i c_{\br-\ba_2+\bdelta_y} + i c_{\br-\ba_2+\bdelta_x}.
\end{align}
\begin{align}\label{eq:T^y-eg}
    [ \scT^y_2 ]_n & = \langle \Omega | o_{\br,y} \left[ H^{\Gamma'}_{\br,y} + H^{\Gamma'}_{\br+\ba_1,x} \right. \nonumber \\
    & \left. H^{\Gamma'}_{\br+\ba_1-\ba_2,x} + H^{\Gamma'}_{\br+\ba_1-\ba_2,y}  \right] d_{\br,x,n}^\dagger | \Omega \rangle \nonumber \\
    & = \Gamma' \langle 0; \Psi_c(\br+\ba_1,x;\br,z)| g \alpha_{(\br,x),n}^\dagger |0; \Psi_c(\br,x) \rangle,\nonumber \\
    g & = c_{\br+\ba_1} - c_{\br} - i c_{\br+\ba_1-\ba_2+\bdelta_y} + i c_{\br-\ba_2+\bdelta_y}.
\end{align}
Similar to the fermion vison pairs, the $\Gamma$-induced boson hopping is much stronger in the FM Kitaev model than in the AFM Kitaev model,
making the boson band gap closes at much smaller $\Gamma$ values in the FM case.
On the other hand, the $\Gamma'$-induced boson hopping is comparable in both cases.

Once the boson gap closes at a momentum $\bk$, the resulting condensate may induce a specific ordered state,
whose symmetry is determined by both $\bk$ and the internal wave function of the soft mode.
In particular, the component of the soft mode on NN vison pairs determines its overlap with local spin operators
and hence the associated magnetic long-range order.
Time-reversal symmetry provides an additional constraint on the possible ordered states.
Since time reversal sends $\bk\rightarrow-\bk$, condensation at a non-time-reversal-invariant momentum generally breaks time-reversal symmetry. 
At a time-reversal-invariant momentum, however, the time-reversal character of the condensed bosonic mode depends on its internal wave function,
including its bond-flavor and matter-fermion components.
A complete time-reversal classification would therefore require tracking this internal structure explicitly.
In this work, we instead identify the resulting ordering tendency by directly computing the order induced by the condensed boson wave function.

For purely NNN vison-pair modes, the situation is different.
Since a local Pauli operator acting on the KSL ground state creates a NN vison pair rather than a NNN vison pair,
purely NNN vison-pair modes have no direct overlap with local spin operators.
Their condensation therefore does not immediately generate a local magnetic moment, 
but can instead induce more extended composite orders involving multi-spin or bond operators,
such as valence-bond-crystal or nematic orders.
As an example, we will show below that when the FM KSL is destabilized by positive $\Gamma$,
the purely NNN vison-pair mode at $\bk=\Gamma$ carries nonzero internal orbital angular momentum,
and its softening can be associated with a bond-nematic order characterized by anisotropic bond energies.
%

\begin{figure*}
\centering
\includegraphics[width=0.75 \textwidth]{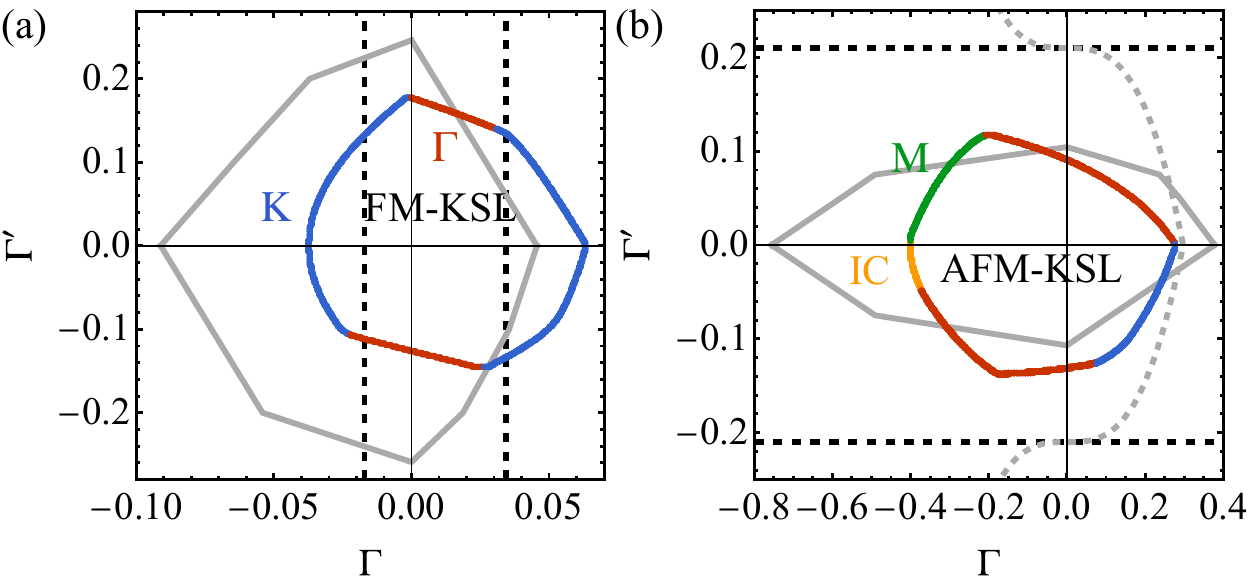}
\caption{
Phase diagram of the $K$-$\Gamma$-$\Gamma'$ model from quasiparticle gap closings.
Results are shown for (a) FM Kitaev coupling, $K=-1$, and (b) AFM Kitaev coupling, $K=1$, 
obtained from a leading-order analysis of the $\Gamma$ and $\Gamma'$ interactions.
Black dashed lines indicate the closing of the single-vison gap, while gray contours mark the closing of the fermion gap.
Colored phase boundaries correspond to the closing of the boson gap; the color indicates the momentum of the soft boson mode:
red for $\Gamma$, blue for $\K$, green for $\M$, and yellow for IC wave vectors $\bQIC$ between the $\Gamma$ and $\M$ points.
In panel (b), the gray dashed line indicates the single-vison gap closing after including a single-vison chemical-potential term 
induced by the $\Gamma$ interaction at third order.
}\label{fig:pd}
\end{figure*}

\subsection*{FM Kitaev model under $\Gamma$ perturbation}
%
As shown in earlier sections, for a FM Kitaev interaction, the $\Gamma$ term induces strong single-vison hopping,
causing the vison gap to close at relatively small $\Gamma$ values.
Interestingly, we have found that the gap closes at an IC wave vector $\bk \approx 0.67 \, \M$
between the $\Gamma$ and $\M$ points when $\Gamma \approx 0.035$,
and at the $\Gamma$ and $\K$ points when $\Gamma \approx -0.02$ (see \Fref{fig:sv-band} and \Fref{fig:pd}(a)). 
These vison gap closings signal phase transitions into new phases driven by vison proliferation.
However, the nature of the emergent phases is difficult to determine directly.
This difficulty stems from several factors.
First, since the model is gapless, the vison is a non-local excitation, 
making it unclear how to assign a self-statistical phase to it.
Moreover, in nearby phases where the matter fermions are gapped by a Haldane mass,
the vison becomes a local excitation but acquires a non-Abelian character.
A microscopic understanding of what phase transitions can be driven by the softening of such
non-Abelian particles remains a major open problem (see, however, Ref.~\cite{Zou2020-ni}
for a field-theoretic proposal of a direct continuous transition from non-Abelian states 
to short-range entangled states).
We expect, however, that generally the closure of the single-vison gap drives confinement of all non-local quasiparticles, leading to the disappearance of topological order.
This can be argued by noting that the vison ``core'' is a boson dressed by the non-local cloud of itinerant fermions~\cite{Chen2022-tr,Chen2018-nq},
and the condensation of this core is responsible for confinement, as it can be more clearly understood in Abelian cases~\cite{Pozo2021-gy}.
This principle is supported by our numerical estimates for the critical $\Gamma$ values at which this occurs,
which agree with previous numerical results for phase transitions into other ordered states~\cite{Sorensen2021-jg,Lee2020-zx}.

We have also found that the fermionic vison pairs acquire significant $\Gamma$-induced hopping, leading to a band gap closing at relatively small $\Gamma$ values.
For positive $\Gamma$, the lowest $\Tilde{\chi}$ band is flat and becomes gapless around $\Gamma \approx 0.05$.
For negative $\Gamma$, the $\Tilde{\chi}$ gap closes at $\Gamma$ and $\K$ points when $\Gamma \approx -0.09$
(see \Frefs{fig:f-band}(a)--(b) and \Fref{fig:pd}(a)). 
Notably, in the absence of $\Gamma'$ interaction, the $\Tilde{\chi}$ and matter $c$
fermions are decoupled, so the $c$-fermion Dirac cones persist and remain gapless along the $\Gamma$ axis.

The bosonic vison pairs also develop strong $\Gamma$-induced hopping, leading to band gap closures at critical $\Gamma$ values
comparable to those of single visons and fermions.
In the absence of $\Gamma'$ interaction, the NN ($d$) and NNN ($o$) vison pairs are decoupled.
Along the $\Gamma$ axis, the lowest-energy band originates purely from $o$ bosons (see \Frefs{fig:b-band}(a)–(b)),
whose gap closing does not immediately trigger local magnetic order, as these local bosons
cannot be created or annihilated by single Pauli operators.
This is also reflected in the absence of magnon-like low-energy dispersion in $S(\omega,\bQ)$ near the transition
(see \Frefs{fig:S}(a)--(b)).

For positive $\Gamma$, the lowest $o$ band is flat (becoming dispersive when $\Gamma'$ is finite)
and its gap closes around $\Gamma \approx 0.06$.
There exists a twofold degeneracy at the $\Gamma$ point, with both modes carrying orbital angular momentum $l = \pm 1$:
\begin{align}\label{eq:o-mode-FM}
    \beta_{\Gamma,\pm1}^\dagger = \frac{1}{\sqrt{3}} \left( 
    o_{\Gamma,x}^\dagger + e^{\pm i 2\pi/3} o_{\Gamma,y}^\dagger + e^{\pm i 4\pi/3} o_{\Gamma,z}^\dagger
    \right).
\end{align}
Because $o_{\br,\alpha}$ boson can be created by the $\Gamma$ interaction on bond $(\br,\alpha)$, i.e.,
$\langle \Omega| ( \sigma_{\br}^\beta \sigma_{\br+\bdelta_\alpha}^\gamma + \sigma_{\br}^\gamma \sigma_{\br+\bdelta_\alpha}^\beta )
o_{\br,\alpha}^\dagger |\Omega \rangle \neq 0$,
the condensation of $\beta_{\Gamma,\pm 1}$ mode produces an additional contribution to the bond energy
$( \Delta E^x, \Delta E^y, \Delta E^z ) \propto \Gamma ( \cos(\theta), \cos(\theta+2\pi/3), \cos(\theta+4\pi/3) )$.
Depending on the Bose condensation phase $\theta$, $E^{\alpha} < E^\beta = E^{\gamma}$ or $E^{\alpha} > E^\beta = E^{\gamma}$ might occur.
Interestingly, such nematic features were indeed observed in numerical studies of the $\Gamma$-induced paramagnetic (NP) phase~\cite{Lee2020-zx,Gohlke2020-go}.
For $\Gamma < 0$, the lowest $o$ band becomes dispersive and its gap closes at the $\Gamma$ and $\mathrm{K}$ points around $\Gamma \approx -0.04$.
In this case, the soft mode at the $\Gamma$ point carries zero orbital angular momentum and does not
induce bond-energy nematicity.

When $\Gamma'$ is finite, the soft mode acquires weight on the NN $d$ bosons, and its condensation can
induce long-range order with an average expectation value of single Pauli operators.
The specific pattern of this order is governed by the momentum and wavefunction of the soft mode.
In the phase diagram \Fref{fig:pd}, the colored phase boundaries indicate the closure of the boson gap, 
with the color denoting the momentum of the soft boson mode: 
red for $\Gamma$, blue for $\K$, green for $\M$, and yellow for IC wave vectors.
Near the $\Gamma$ axis at finite $\Gamma'$, we find that the lowest-energy mode occurs at the $\K$ point,
whose condensation yields a $120^\circ$ order (see its pattern in \Fref{fig:mo}(d)).
After applying the six-site $U_6$ transformation~\cite{Chaloupka2015-uh}, this maps to a uniform FM order,
and is therefore referred to as FM$_{U_6}$~\cite{Rau2014-ao,Sorensen2021-jg,Sorensen_2024_th,Rousochatzakis2024-td}.
For negative $\Gamma$, it is known that the KSL should be replaced by an AFM$_{U_6}$ order, which corresponds to the N\'eel state after
the $U_6$ transformation~\cite{Chaloupka2015-uh,Rau2014-ao,Sorensen2021-jg,Sorensen_2024_th}.
While our analysis finds a lowest-energy boson mode leading to FM$_{U_6}$ order, the second-lowest mode at $\K$ carries
the appropriate structure to induce AFM$_{U_6}$ order.
We thus expect that improvements such as including bosonic states with two-$c$ fermion excitations for NNN vison pairs,
or incorporating higher-order vison-pair hopping processes, might be needed to capture the correct magnetic phase.
Note that without including the NNN vison pairs, the boson gap closes later than the fermion gap~\cite{Zhang2021-xi}. 
This highlights the important role of NNN vison pairs---as the lowest-energy two-vison configuration of the Kitaev model---in the low-energy physics.

\subsection*{FM Kitaev model under $\Gamma'$ perturbation}
To leading order, single visons exhibit vanishing $\Gamma'$-induced hopping and thus remain gapped as $\Gamma'$ increases,
although higher-order terms may enable their hopping and eventually drive a gap closing at finite $\Gamma'$.
While the $\Tilde{\chi}$ and $c$ fermions are coupled by the $\Gamma'$ interaction, the spectrum remains gapless at the $\K$ point.
The fermion gap collapses at the $\Gamma$ point near $\Gamma' \approx 0.24$, and at an IC wave vector around $\Gamma' \approx -0.26$ (see \Frefs{fig:f-band}(c)--(d)).

\begin{figure*}
\centering
\includegraphics[width = 1 \textwidth]{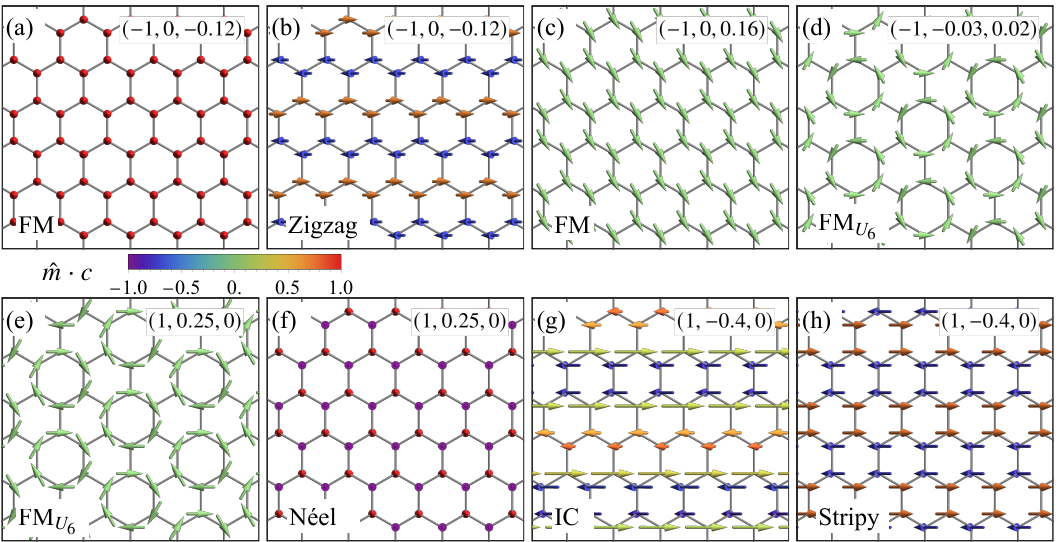}
\caption{
Magnetic order induced by the soft boson modes.
The value on the top-right corner of each figure indicates the corresponding value of $(K,\Gamma,\Gamma')$.
(a)--(d) Representative points for the FM Kitaev model.
(e)--(f) Representative points for the AFM Kitaev model.
The color represents (cosine of) the angle between local moments and the out-of-plane $\bm{c}$ axis.
At $(K,\Gamma,\Gamma') = (-1,0,-0.12)$, the lowest energy mode is at the $\Gamma$ point, 
which induces an out-of-plane FM order, nevertheless, the lowest energy mode at the $\M$ point induces a
zigzag order, as shown in (b).
At $(K,\Gamma,\Gamma') = (1,0.25,0)$, the lowest energy modes at $\Gamma$ and $\K$ points are degenerate (see \Fref{fig:b-band}(f)).
The $\mathrm{K}$ mode induces an in-plane $120^\circ$ order, while the $\Gamma$ mode leads to an out-of-plane N\'eel order,
which, under the $U_6$ transformation, maps to an out-of-plane FM order---also consistent with the FM$_{U_6}$ phase.
}\label{fig:mo}
\end{figure*}

\begin{figure*}
\centering
\includegraphics[width = 0.85 \textwidth]{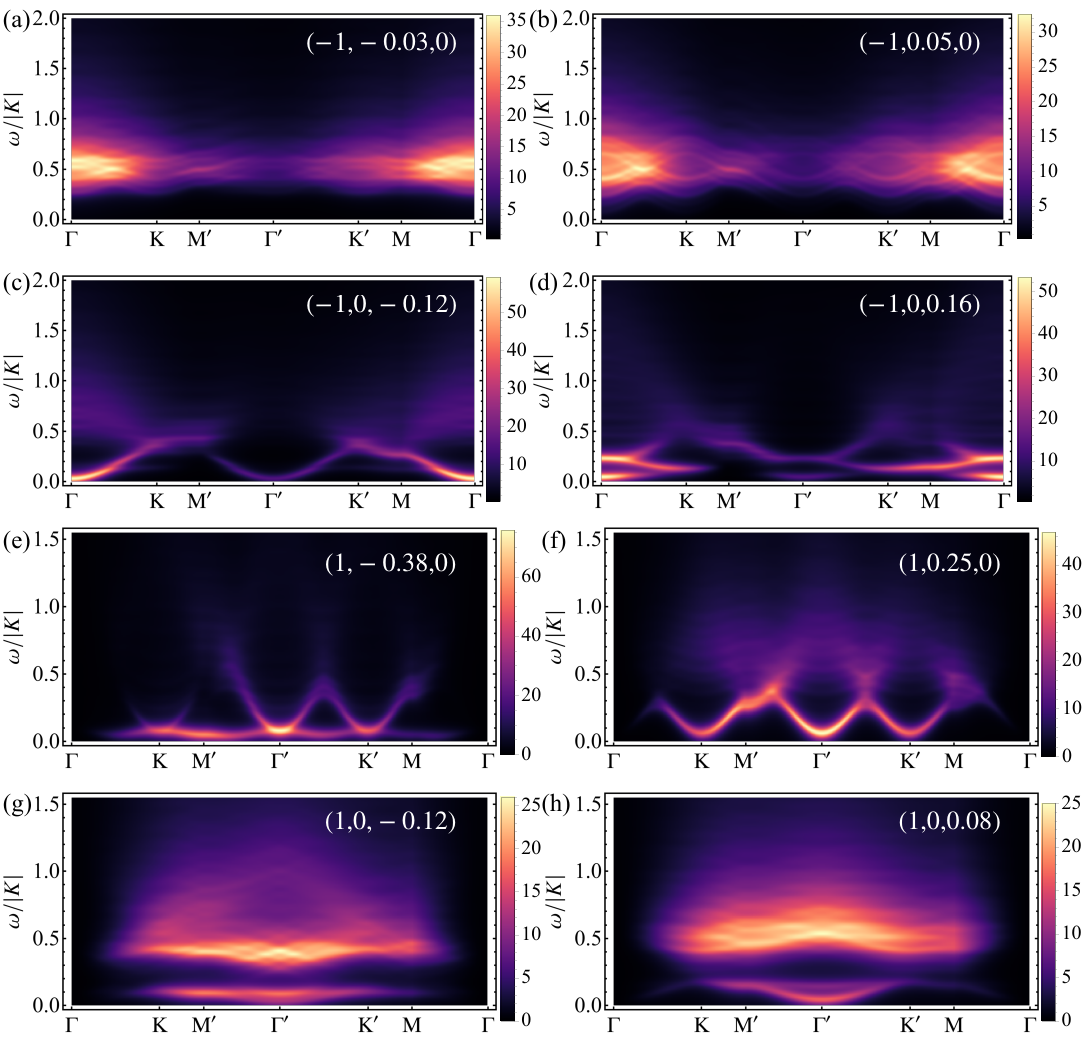}
\caption{
Dynamical spin structure factor computed from bosonic vison-pair modes.
(a)--(d) Dynamical spin structure factor $S(\omega,\bQ)$ evaluated at selected perturbation strengths for ferromagnetic (FM) Kitaev interactions.
(e)--(h) Corresponding results for antiferromagnetic (AFM) Kitaev interactions.
The calculations use bosonic vison-pair modes and include a Lorentzian broadening factor $\eta=0.04|K|$.
The momentum path follows that shown in \Fref{fig:schematic}(b).
}\label{fig:S}
\end{figure*}

The boson band minimum lies at the $\Gamma$ point for both positive and negative $\Gamma'$.
Around $\Gamma' \approx 0.17$, we find a FM order induced by boson condensation (see its pattern in \Fref{fig:mo}(c) and $S(\omega,\bQ)$ in \Fref{fig:S}(d)),
consistent with variational Monte Carlo results ($\Gamma'_c \approx 0.12$ in Ref.~\cite{Wang2024-yg}).
For $\Gamma' < 0$, previous numerical studies have shown that the KSL is replaced by a zigzag order
at $\Gamma' \approx -0.1$~\cite{Gordon2019-iu,Lee2020-zx,Gohlke2020-go,Wang2024-yg},
while our boson band calculations (see \Fref{fig:b-band}(c)) predict a gap closing near $\Gamma' \approx -0.12$.
However, the lowest bosonic mode in our calculation occurs at the $\Gamma$ point and induces the out-of-plane FM order shown in \Fref{fig:mo}(a),
rather than the zigzag order found numerically.
The boson mode at the $\M$ point, which corresponds to the zigzag wave vector, remains slightly higher in energy;
nevertheless, if this mode condenses, it produces a magnetic pattern that precisely matches the zigzag order (see \Fref{fig:mo}(b)).
Furthermore, the induced moment tilts at an angle of about $45^\circ$ from the honeycomb plane,
close to the experimental value ($35^\circ$) observed in $\alpha$-RuCl$_3$~\cite{Sears2020-si}.
These observations suggest that bosonic vison-pair fluctuations near the $\M$ point are closely related to the emergence of zigzag order in Kitaev materials. 

At the same time, the vertical dashed lines indicating the single-vison instability in \cref{fig:pd}(a) should be interpreted with caution,
since they result from retaining only the leading-order single-vison hopping.
Higher-order single-vison processes are expected to bend these lines and may reduce the apparent separation from the bosonic vison-pair instability 
at larger $|\Gamma'|$.
Moreover, the fact that the lowest bosonic mode in our present calculation occurs at the $\Gamma$ point,
rather than at the $\M$ point associated with zigzag order,
indicates that the low-order vison-pair description does not fully capture the energetics in this regime.
Effects beyond the present treatment, such as higher-order single-vison hopping,
vison-pair dissociation through hybridization with separated-vison states,
quasiparticle interactions, or other nonperturbative effects near the transition,
may be needed to correctly select the zigzag instability.
%

\subsection*{AFM Kitaev model under $\Gamma$ perturbation}
%
In the AFM Kitaev model, single visons do not acquire $\Gamma$-induced hopping and thus remain gapped as $\Gamma$ increases.
The fermions also have a weaker hopping, and their gap closes at much larger $\Gamma$ values (around $\Gamma \approx 0.38$ and
$\Gamma \approx -0.8$) compared to the FM case.

On the other hand, the boson gap closes more rapidly than those of single visons and fermions.
Unlike the FM case, where the lowest band along the $\Gamma$ axis originates from NNN vison pairs,
the lowest band here is purely from NN vison pairs, whose condensation leads directly 
to finite expectation values of the local Pauli operators with
magnetic long-range order.
At $\Gamma \approx 0.28$, the boson gap closes simultaneously at the $\Gamma$ and $\K$ points (see \Fref{fig:b-band}(f)).
We find that the soft mode at $\K$ point induces an in-plane FM$_{U_6}$ order, as shown in \Fref{fig:mo}(e), while the $\Gamma$ mode
produces an out-of-plane N\'eel order (see \Fref{fig:mo}(f)), which maps to an out-of-plane FM order under the $U_6$ transformation.
Therefore, our results are fully consistent with the expected FM$_{U_6}$ order~\cite{Chaloupka2015-uh,Sorensen2021-jg,Rousochatzakis2024-td}.
The dynamical structure factor $S(\omega,\bQ)$ near the transition point, shown in \Fref{fig:S}(f), exhibits a magnon-like low-energy
dispersion touching zero at both the $\K$ and $\Gamma'$ points.
Notably, it closely resembles results from recent random-phase approximation calculations~\cite{Rao2025-pp}.
In the presence of $\Gamma'$ interaction, the $\Gamma$–$\K$ degeneracy is lifted: 
a $\Gamma$-mode-driven N\'eel order emerges for $\Gamma' > 0$, while a $\K$-mode-driven FM$_{U_6}$ order appears for $\Gamma' < 0$ (see \Fref{fig:pd}(b)).
Interestingly, a similar trend has also been reported in a vison-pair analysis of the $K$–$J$–$\Gamma$ model, with $\Gamma’$ replaced by $J$~\cite{Zhang2021-xi}.

Around $\Gamma \approx -0.4$, we find that the boson band touches zero at an IC wave vector $\bQIC$ located between the
$\Gamma$ and $\M$ points (see \Fref{fig:b-band}(e)), therefore the condensation of this mode is expected to
induce an IC spiral order (see its pattern in \Fref{fig:mo}(g) and dynamical structure factor in \Fref{fig:S}(e)).
Moreover, a nearly gapless mode also appears at the $\M$ point, whose condensation leads to a stripy order,
as shown in \Fref{fig:mo}(h).
The close proximity in energy between this mode and the IC mode indicates a strong competition between the two orders---consistent with recent numerical studies,
where both types of order emerge depending on system size, and are collectively referred to as a ``spin nematic'' phase~\cite{Sorensen2021-jg}.
Upon introducing a small positive $\Gamma'$, the $\M$ mode softens first, favoring a stripy phase; conversely, 
for negative $\Gamma'$, the IC mode becomes soft first, leading to an IC spiral order where the value of $\bQIC$ evolves with $\Gamma'$.
In agreement with our findings, the IC order has also been obtained in a vison-pair analysis of the $K$-$J$-$\Gamma$ model with
AFM Kitaev coupling and $\Gamma < 0$~\cite{Zhang2021-xi}.
Moreover, previous numerical studies have reported analogous behavior when $\Gamma'$ is replaced by a Heisenberg coupling:
the system exhibits stripy order for $J > 0$ and IC spiral order $J < 0$~\cite{Rau2014-ao}.

It is worth noting that the condensation of bosonic vison pairs does not immediately lead to spinon confinement,
as they are local bosons whose statistical interactions with matter fermions are trivial.
Therefore, since we have found that single visons remain gapped due to 
the absence of $\Gamma$-induced hopping, the KSL is likely to evolve into
a magnetically ordered QSL before undergoing a transition to a conventional ordered phase at larger $\Gamma$ values,
where single-vison condensation occurs.

To further investigate this, we analyze the $\Gamma$-induced single-vison hopping at second order.
We find that it also vanishes exactly in the AFM Kitaev model, suggesting that single visons are unlikely to condense via enhancing its mobility.
However, we have found that at third order, a $\Gamma$-induced chemical potential term arises:
$\propto -\Gamma^3 \sum_i n_i^{\mathrm{v}}$ (see details in the Supplementary Information).
This term favors (disfavors) single-vison excitations for $\Gamma > 0$ ($\Gamma < 0$).
Including this term leads to single-vison gap closing for $\Gamma > 0$, while they remain gapped for $\Gamma < 0$
(see the gray dashed line in \Fref{fig:pd}(b)).
This analysis provides additional evidence that the FM$_{U_6}$ order observed for $\Gamma > 0$ corresponds to a conventional (confined) magnetic order,
whereas the IC phase at $\Gamma < 0$ may represent a magnetically ordered yet genuine QSL.
Although no direct numerical evidence for the QSL has been reported, previous DMRG studies on a two-leg ladder
identified this regime as a line of first-order transitions between stripy and IC phases,
leaving open the possibility that such an exotic QSL could emerge in the two-dimensional limit~\cite{Sorensen2021-jg}.

\subsection*{AFM Kitaev model under $\Gamma'$ perturbation}
Different from the $\Gamma$ interaction, the $\Gamma'$ interaction induces a finite single-vison hopping.
Similar to the case of a Zeeman field, the $\Gamma'$-induced single vison hopping has a fractionalized translational symmetry,
resulting in two single-vison bands with opposite Chern number~\cite{Chen2023-vo,Joy2022-ba}.
The single-vison gap closes around $\Gamma' \approx \pm 0.21$.
The fermionic vison bands are also dispersive (see \Fref{fig:f-band}), with the gap closing first at the 
$\M$ point for $\Gamma' \approx 0.1$ and at the $\Gamma$ point for $\Gamma' \approx -0.1$.

At $\Gamma' \approx 0.09$ and $\Gamma' \approx -0.13$, the bosonic vison pairs close gap at the $\Gamma$ point.
The condensation of the soft boson results in an out-of-plane N\'eel order, similar to the one shown in \Fref{fig:mo}(f).
The corresponding dynamical structure factors near both transitions exhibit similar features
---a magnon-like low-energy band and a mushroom-shaped high-energy continuum (see \Frefs{fig:S}(g) and (h))---which could be examined in future numerical studies.

\section*{Discussion}
We have investigated the $K$-$\Gamma$-$\Gamma'$ model from the perspective of the KSL quasiparticles,
treating the non-Kitaev $\Gamma$ and $\Gamma'$ interactions as perturbations and analyzing their effects on the 
quasiparticle dispersions.
The goal of this quasiparticle-based approach is to compute, in a perturbatively controlled manner near the ideal Kitaev limit,
the dispersions of all anyonic quasiparticles, and to extrapolate from them to identify the plausible dominant quasiparticle instabilities 
and the associated ordering tendencies near the KSL.
We caution that our approach relies on leading-order perturbative calculations of the anyon hopping amplitudes, 
and that other effects arising at higher orders in perturbation theory, i.e., at higher orders in $\Gamma$ or $\Gamma'$, 
such as anyon-anyon interactions, are neglected in the present treatment.
Nevertheless, this approach has previously been applied to other extended Kitaev models, including the $K$-$J$-$\Gamma$
model~\cite{Zhang2021-xi}, Kitaev model with external magnetic fields~\cite{Joy2022-ba,Chen2023-vo,Chen2025-be},
and multilayer Kitaev systems~\cite{Joy2024-cx}, offering valuable insights into their physical properties
and into the competition between the KSL and magnetically ordered phases.
%

The stronger (weaker) $\Gamma$-induced hopping of single visons and vison pairs in the FM (AFM) Kitaev model naturally explains
the weaker (stronger) stability of the KSL under $\Gamma$ perturbations, as reflected in the phase diagram shown in \Fref{fig:pd}.
This suggests that the suppression of the KSL is driven by vison proliferation.

The softening of bosonic vison pairs provides a compelling mechanism for the emergence of various magnetic orders, which we have been able to infer by analyzing the momentum and other quantum numbers of the wavefunction of softened bosonic modes.
For example, in the AFM Kitaev model, the FM$_{U_6}$ order at $\Gamma > 0$ arises from the condensation of the $\K$ mode 
or $\Gamma$ mode (see \Frefs{fig:mo}(e)--(f) and \Fref{fig:S}(f)); the IC spiral order and the stripy orders at $\Gamma < 0$
are driven by softened modes at $\bQIC$ and $\M$ points (see \Frefs{fig:mo}(g)--(h) and \Fref{fig:S}(e)), respectively.
In the FM Kitaev model with $\Gamma' < 0$, although the $\M$ mode---which induces a zigzag order consistent with
the moment direction observed in $\alpha$-RuCl$_3$---is only the second lowest-energy mode (see \Fref{fig:b-band}(c) and
\Fref{fig:mo}(b)), its energy might be lowered by extending the vison-pair subspace, possibly allowing it to become the softest mode.

The NNN vison pairs ($o$) also play an essential role at low energies. 
In the FM Kitaev model, they form the lowest bosonic band along the $\Gamma$ axis
in the $\Gamma$-$\Gamma'$ phase diagram (see \Fref{fig:pd}(a)). 
For $\Gamma > 0$, the softened mode at $\Gamma$ point carries finite orbital angular momentum and induces a paramagnetic phase with anisotropic bond energies, 
offering an explanation for the nematic paramagnetic phase observed in previous numerical studies~\cite{Lee2020-zx,Gohlke2020-go}.

One of our main finding is that, in the AFM Kitaev model with $\Gamma < 0$, bosonic vison pairs soften while single visons remain gapped (see \Fref{fig:pd}(b)).
This raises the tantalizing possibility that the regime $\Gamma \lesssim -0.4|K|$ realizes
an intermediate phase in which long-range magnetic order coexists with fractionalized excitations
inherited from the parent KSL, before the system eventually transitions into a trivial magnetically ordered phase
at larger $|\Gamma|$.
%
%
The symmetry-breaking magnetic order is likely either stripy or a IC spiral (see \Frefs{fig:mo}(g)--(h)).
Consistently, the IC order has also been reported in a previous vison-pair study of the $K$-$J$-$\Gamma$ model~\cite{Zhang2021-xi}.
The coexistence of fractionalization and a spiraling magnetic moment is particularly intriguing,
as the IC moment may remain strictly gapless.
We hope that future numerical and analytical studies will provide a more direct assessment of this fascinating possibility.
%
%

We have also demonstrated how correlation functions of local observables can be directly related to the bosonic vison pairs, and used these modes to explicitly compute the dynamical structure factor.
Near phase transitions into magnetically ordered states, the dynamical structure factor exhibits low-energy magnon-like dispersions along with broad high-energy continua.
Notably, in some cases, our results resemble those obtained from random-phase approximation studies
(cf. \Fref{fig:S}(f) and Fig.~9(c) of Ref.~\cite{Rao2025-pp}).
Such features serve as valuable signatures for probing emergent phases in Kitaev materials. 
This framework can also be extended to study other experimental probes, such as Raman spectroscopy~\cite{Joy2025-pk},
providing a promising direction for future investigations.

\begin{figure}[t!]
\centering
\includegraphics[width = 0.46 \textwidth]{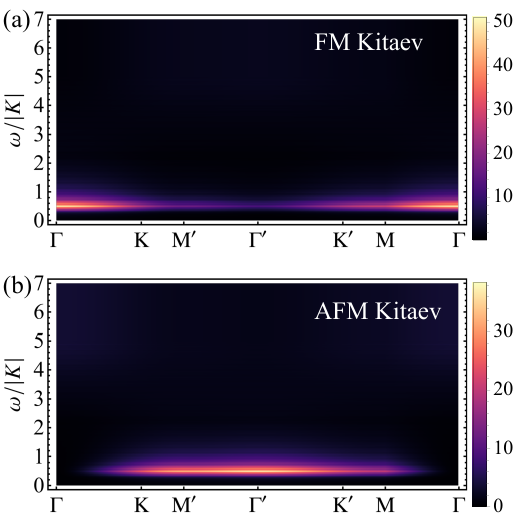}
\caption{
Dynamical spin structure factor of the ideal Kitaev model approximated by bosonic vison-pair contributions.
Results are shown for (a) the FM and (b) AFM Kitaev models.
}\label{fig:S-Kitaev}
\end{figure}

\section*{Methods}\label{sec:method}
\subsection*{Magnetic orders from the condensation of bosonic vison pairs}\label{sec:mo-boson}
The condensation of a bosonic mode can give rise to magnetic long-range order,
provided it carries weight on NN vison pairs (see also Sec.~V of Ref.~\cite{Zhang2021-xi}).
Consider a soft boson mode $\beta_{\bq}$ with momentum $\bq$:
\begin{align}
    \beta_{\bq}^\dagger & = d_{\bq,x}^{\dagger} w_{x}(\bq) +  d_{\bq,y}^{\dagger}w_{y}(\bq) + d_{\bq,z}^{\dagger} w_{z}(\bq) \nonumber \\
    & + o_{\bq,x}^{\dagger} v_{x}(\bq) + o_{\bq,y}^{\dagger} v_{y}(\bq) + o_{\bq,z}^{\dagger} v_{z}(\bq).
\end{align}
where the $N \times 1$ vector $w_{\alpha}(\bq)$ is its weight on NN $d_{\bq,\alpha}$ vison pairs
and $v_{\alpha}(q)$ is its weight on NNN $o_{\bq,\alpha}$ vison pairs.
Note $d_{\bq,\alpha}^\dagger = ( d_{\bq,\alpha,1}^\dagger, \dots, d_{\bq,\alpha,N}^\dagger )$ (see \Eqref{eq:d-compact}),
and
\begin{equation}
    d_{\bq,\alpha,l}^\dagger = \frac{1}{\sqrt{N}} \sum_{\br \in A} e^{i \bq \cdot \br} d_{\br,\alpha,l}^\dagger.
\end{equation}
The local magnetic moment induced by the condensation of $\beta_{\bq}$ mode reads:
\begin{align}
    \langle \sigma_{\br}^{\alpha} \rangle \propto e^{i \theta} \langle \Omega| \sigma_{\br}^\alpha \beta_{\bq}^\dagger |\Omega \rangle + \mathrm{H.c.}
\end{align}
Here $\theta$ is the condensation phase of the soft mode.
It can be shown that, for $\br \in A$,
\begin{align}\label{eq:order-VP}
    \langle \Omega| \sigma_{\br}^\alpha \beta_{\bq}^\dagger |\Omega \rangle 
    & = \frac{1}{\sqrt{N}} e^{i \bq \cdot \br} \langle \Omega | \sigma_{\br}^{\alpha} d_{\br,\alpha}^\dagger | \Omega \rangle w_{\alpha}(\bq) \nonumber \\
    & = \frac{1}{\sqrt{N}} e^{i \bq \cdot \br} P_{\alpha,A} w_{\alpha}(\bq).
\end{align}
Here
\begin{align}\label{eq:P_A}
    P_{\alpha,A} = \langle \Omega |\sigma_{\br}^\alpha (  d_{\br,\alpha,1}^\dagger, \dots, d_{\br,\alpha,N}^\dagger ) | \Omega \rangle.
\end{align}
Similarly, for $\br' \in B$, there is
\begin{align}\label{eq:order-VP}
    \langle \Omega| \sigma_{\br'}^\alpha \beta_{\bq}^\dagger |\Omega \rangle 
    & = \frac{1}{\sqrt{N}} e^{i \bq \cdot (\br'-\bdelta_{\alpha})} \langle \Omega | \sigma_{\br'}^{\alpha} d_{\br'-\bdelta_{\alpha},\alpha}^\dagger | \Omega \rangle w_{\alpha}(\bq) \nonumber \\
    & = \frac{1}{\sqrt{N}} e^{i \bq \cdot (\br'-\bdelta_{\alpha})} P_{\alpha,B} w_{\alpha}(\bq),
\end{align}
with
\begin{align}\label{eq:p_B}
    P_{\alpha,B} = \langle \Omega |\sigma_{\br'}^\alpha ( d_{\br'-\bdelta_{\alpha},\alpha,1}^\dagger, \dots, d_{\br'-\bdelta_{\alpha},\alpha,N}^\dagger) | \Omega \rangle.
\end{align}
Therefore, once $w_{\alpha}(\bq)$ and $P_{A/B,\alpha}$ are determined, the resulting local moments can be obtained.
In our calculations, we find that for orders induced by the $\Gamma$ or $\M$ modes---such as FM, N\'eel, zigzag, and stripy---the direction of
the local moments is independent of $\theta$. In contrast, for the $120^{\circ}$ order induced by the $\K$ mode,
the local moment direction depends on $\theta$, consistent with Ref.~\cite{Zhang2021-xi}.

\subsection*{Dynamical structure factor from the bosonic vison pairs}\label{sec:S-boson}
In this section, we discuss the dynamical structure factor approximated from the bosonic vison pairs, 
a method first used in Ref.~\cite{Zhang2021-xi}.
The dynamical structure factor is defined as
\begin{align}\label{eq:S-def}
    S(\omega,\bQ) & = \int dt \, e^{i \omega t} S(t,\bQ), \nonumber \\
    S(t,\bQ) & = \frac{1}{N_s} \sum_{\alpha} \langle \mathrm{GS} | \sigma_{\bQ}^\alpha(t) \sigma_{-\bQ}^{\alpha}(0) | \mathrm{GS} \rangle,
\end{align}
where $N_s = 2N$ denotes the total number of sites in the system. The Fourier tranform of Pauli operators is given by
\begin{align}
    \sigma_{\bQ}^\alpha = \sum_{\br \in A} e^{-i \bQ \cdot \br} \sigma_{\br}^\alpha 
    + e^{-i \bQ \cdot (\br+\bdelta_{\alpha})} \sigma_{\br+\bdelta_{\alpha}}^\alpha.
\end{align}
Note that due to the bipartitle nature of the honeycomb lattice, $\sigma_{\bQ}^\alpha \neq \sigma_{\bQ+\bb}^\alpha$ with $\bb$ being 
a reciprocal lattice vector.
As a result, it is helpful to evaluate $S(\omega,\bQ)$ in the extended BZ.

$| \mathrm{GS} \rangle = | \Omega \rangle$ before the breakdown of KSL. 
Using the Lehmann representation, \Eqref{eq:S-def} can be recast into:
\begin{align}
    S(t,\bQ) = \frac{1}{N_s} \sum_{l} e^{-i(E_l-E_0) t} | \langle \Omega | \sigma_{\bQ}^\alpha | l \rangle |^2,
\end{align}
with $E_l-E_0$ being the excitation energy of $l$-th excited state.
Since a local Pauli operator always excites a single bosonic vison pair from the ground state $|\Omega\rangle$~\cite{Knolle2014-wl,Knolle2015-lx},
one can choose $|l\rangle$ to be the single-boson eigenstate $\beta_{\bQ,l}^\dagger |\Omega \rangle$, 
where $\beta_{\bQ,l}^\dagger$ creates the $l$-th eigenmode with momentum $\bQ$ of the boson Hamiltonian.
Note that for $\bQ$ outside the first BZ, one should use $\beta_{[\bQ],l}^\dagger |\Omega \rangle$, 
where $[\bQ]$ denotes the corresponding wave vector folded back into the first BZ.
After some algebra, one can show that:
\begin{align}
    \langle \Omega | \sigma_{\bQ}^\alpha \beta_{\bQ,l}^\dagger | \Omega \rangle 
    = \sqrt{N} ( P_{\alpha,A} + e^{-i \bQ \cdot \bdelta_\alpha} P_{\alpha,B} ) w_{\alpha}(\bQ).
\end{align}
The $S(\omega,\bQ)$ thus reads:
\begin{align}
    S(\omega,\bQ) & = \pi \sum_{l} \delta(\omega - \Delta E_{Q,l}) \nonumber \\
    & \times | ( P_{\alpha,A} + e^{-i \bQ \cdot \bdelta_\alpha} P_{\alpha,B} ) w_{\alpha}(\bQ) |^2.
\end{align}
The Dirac delta function is approximated by a Lorentzian: $\delta(x) \approx \frac{\eta/\pi}{\omega^2 +\eta^2}$,
where we take $\eta = 0.04|K|$ in our calculation.
\Fref{fig:S-Kitaev} presents the dynamical structure factor $S(\omega,\bQ)$ of the ideal Kitaev model, computed from bosonic vison pairs.
The results show good agreement with the exact solution of the Kitaev model~\cite{Knolle2014-wl,Knolle2015-lx}.

\acknowledgements
We thank Chao-Ming Jian, Aprem Joy, Peng Rao, and Jiucai Wang for helpful discussions.
C.C. is grateful to Shang-Shun Zhang for several insightful discussions.
We thank Beijing Paratera Co., Ltd. for providing HPC resources that contributed to the numerical results reported in this paper.
C.C. acknowledges support from the National Natural Science Foundation of China (Grants No.~12404175 and No.~12247101),
the Fundamental Research Funds for the Central Universities (Grant No.~lzujbky-2025-jdzx07), 
the Natural Science Foundation of Gansu Province (No.~22JR5RA389 and No.~25JRRA799),
and the ‘111 Center’ under Grant No.~B20063.
I.S.V. acknowledges support from the Deutsche Forschungsgemeinschaft (DFG) under research Grants No.~542614019, No.~518372354, and No.~555335098.
The funders had no role in the study design, analysis, interpretation of results, or preparation of the manuscript.

\section*{Author contributions}
C.C. and I.S.V. designed the study.
C.C. carried out the calculations with input from I.S.V.
C.C. and I.S.V. wrote the manuscript together.

\section*{Data availability}
The datasets generated and/or analyzed during the current study are not publicly available because they consist of supporting numerical data 
generated from the theoretical calculations described in the manuscript and have not been organized as a standalone public dataset,
but are available from the corresponding author on reasonable request.
%

\section*{Code availability}
The codes used during the current study will be available from the corresponding author
upon reasonable request.

\section*{Competing interests}
The authors declare no competing financial or non-financial interests.

\section*{References}
\bibliography{reference.bib}

@ARTICLE{Kitaev2006,
  title     = "Anyons in an exactly solved model and beyond",
  author    = "Kitaev, Alexei",
  journal   = "Ann. Phys. (N. Y.)",
  publisher = "Elsevier BV",
  volume    =  321,
  number    =  1,
  pages     = "2--111",
  month     =  jan,
  year      =  2006,
  url       = "http://dx.doi.org/10.1016/j.aop.2005.10.005",
  doi       = "10.1016/j.aop.2005.10.005",
  issn      = "0003-4916,1096-035X"
}

@article{Broholm2020-pq,
  title = {Quantum Spin Liquids},
  author = {Broholm, C. and Cava, R. J. and Kivelson, S. A. and Nocera, D. G. and Norman, M. R. and Senthil, T.},
  year = 2020,
  month = jan,
  journal = {Science},
  volume = {367},
  number = {6475},
  pages = {eaay0668},
  issn = {0036-8075, 1095-9203},
  doi = {10.1126/science.aay0668},
  url = "http://dx.doi.org/10.1126/science.aay0668"
}

@ARTICLE{Balents2010-td,
  title   = "Spin liquids in frustrated magnets",
  author  = "Balents, Leon",
  journal = "Nature",
  volume  =  464,
  number  =  7286,
  pages   = "199--208",
  month   =  "11~" # mar,
  year    =  2010,
  url     = "http://dx.doi.org/10.1038/nature08917",
  doi     = "10.1038/nature08917",
  pmid    =  20220838,
  issn    = "0028-0836,1476-4687"
}

@book{Fradkin_2013,
place={Cambridge},
edition={2},
title={Field Theories of Condensed Matter Physics},
publisher={Cambridge University Press},
author={Fradkin, Eduardo},
year={2013}
}

@book{Wen2004,
  title = {{Quantum Field Theory of Many-body Systems: From the Origin of Sound to an Origin of Light and Electrons}},
  author={Wen, Xiao-Gang},
  year={2004},
  publisher={Oxford University Press}
}

@ARTICLE{Jackeli2009-yn,
  title   = "Mott insulators in the strong spin-orbit coupling limit: from
             Heisenberg to a quantum compass and Kitaev models",
  author  = "Jackeli, G and Khaliullin, G",
  journal = "Phys. Rev. Lett.",
  volume  =  102,
  number  =  1,
  pages   =  017205,
  month   =  "9~" # jan,
  year    =  2009,
  url     = "http://dx.doi.org/10.1103/PhysRevLett.102.017205",
  doi     = "10.1103/PhysRevLett.102.017205",
  pmid    =  19257237,
  issn    = "0031-9007"
}

@ARTICLE{Moessner2001-mz,
  title     = "Resonating Valence Bond Phase in the Triangular Lattice Quantum
               Dimer Model",
  author    = "Moessner, R and Sondhi, S L",
  journal   = "Phys. Rev. Lett.",
  publisher = "American Physical Society",
  volume    =  86,
  number    =  9,
  pages     = "1881--1884",
  month     =  "26~" # feb,
  year      =  2001,
  url       = "https://link.aps.org/doi/10.1103/PhysRevLett.86.1881",
  doi       = "10.1103/PhysRevLett.86.1881",
  issn      = "0031-9007"
}

@ARTICLE{Rokhsar1988-el,
  title   = "Superconductivity and the quantum hard-core dimer gas",
  author  = "Rokhsar, D S and Kivelson, S A",
  journal = "Phys. Rev. Lett.",
  volume  =  61,
  number  =  20,
  pages   = "2376--2379",
  month   =  "14~" # nov,
  year    =  1988,
  url     = "http://dx.doi.org/10.1103/PhysRevLett.61.2376",
  doi     = "10.1103/PhysRevLett.61.2376",
  pmid    =  10039096,
  issn    = "0031-9007,1079-7114"
}

@ARTICLE{Kitaev2003-ss,
  title     = "Fault-tolerant quantum computation by anyons",
  author    = "Kitaev, A Yu",
  journal   = "Ann. Phys. (N. Y.)",
  publisher = "Elsevier BV",
  volume    =  303,
  number    =  1,
  pages     = "2--30",
  month     =  "1~" # jan,
  year      =  2003,
  url       = "http://dx.doi.org/10.1016/S0003-4916(02)00018-0",
  doi       = "10.1016/s0003-4916(02)00018-0",
  issn      = "0003-4916,1096-035X"
}

@article{Levin2005,
  title = {String-net condensation: A physical mechanism for topological phases},
  author = {Levin, Michael A. and Wen, Xiao-Gang},
  journal = {Phys. Rev. B},
  volume = {71},
  issue = {4},
  pages = {045110},
  numpages = {21},
  year = {2005},
  month = {Jan},
  publisher = {American Physical Society},
  doi = {10.1103/PhysRevB.71.045110},
  url = {https://link.aps.org/doi/10.1103/PhysRevB.71.045110}
}

@article{Matsuda2025-hs,
  title = {Kitaev quantum spin liquids},
  author = {Matsuda, Yuji and Shibauchi, Takasada and Kee, Hae-Young},
  journal = {Rev. Mod. Phys.},
  volume = {97},
  issue = {4},
  pages = {045003},
  numpages = {61},
  year = {2025},
  month = {Dec},
  publisher = {American Physical Society},
  doi = {10.1103/3m4m-3v59},
  url = {https://link.aps.org/doi/10.1103/3m4m-3v59}
}

@ARTICLE{Trebst2022-go,
  title     = "Kitaev materials",
  author    = "Trebst, Simon and Hickey, Ciarán",
  journal   = "Phys. Rep.",
  publisher = "North-Holland",
  volume    =  950,
  pages     = "1--37",
  month     =  "4~" # mar,
  year      =  2022,
  url       = "https://www.sciencedirect.com/science/article/pii/S0370157321004051",
  doi       = "10.1016/j.physrep.2021.11.003",
  issn      = "0370-1573"
}

@ARTICLE{Takagi2019-ok,
  title     = "Concept and realization of Kitaev quantum spin liquids",
  author    = "Takagi, Hidenori and Takayama, Tomohiro and Jackeli, George and
               Khaliullin, Giniyat and Nagler, Stephen E",
  journal   = "Nat. Rev. Phys.",
  publisher = "Nature Publishing Group",
  volume    =  1,
  number    =  4,
  pages     = "264--280",
  month     =  "25~" # mar,
  year      =  2019,
  url       = "https://www.nature.com/articles/s42254-019-0038-2",
  doi       = "10.1038/s42254-019-0038-2",
  issn      = "2522-5820,2522-5820"
}

@ARTICLE{Rau2014-ao,
  title   = "Generic spin model for the honeycomb iridates beyond the Kitaev
             limit",
  author  = "Rau, Jeffrey G and Lee, Eric Kin-Ho and Kee, Hae-Young",
  journal = "Phys. Rev. Lett.",
  volume  =  112,
  number  =  7,
  pages   =  077204,
  month   =  "21~" # feb,
  year    =  2014,
  url     = "http://dx.doi.org/10.1103/PhysRevLett.112.077204",
  doi     = "10.1103/PhysRevLett.112.077204",
  pmid    =  24579632,
  issn    = "0031-9007,1079-7114"
}

@ARTICLE{Rau2014-ns,
  title         = "Trigonal distortion in the honeycomb iridates: Proximity of
                   zigzag and spiral phases in {Na$_2$IrO$_3$}",
  author        = "Rau, Jeffrey G and Kee, Hae-Young",
  journal       = "arXiv [cond-mat.str-el]",
  month         =  "20~" # aug,
  year          =  2014,
  url           = "http://arxiv.org/abs/1408.4811",
  archivePrefix = "arXiv",
  eprint        = "1408.4811"
}

@ARTICLE{Rau2016-oa,
  title     = "Spin-Orbit Physics Giving Rise to Novel Phases in Correlated
               Systems: Iridates and Related Materials",
  author    = "Rau, Jeffrey G and Lee, Eric Kin-Ho and Kee, Hae-Young",
  journal   = "Annu. Rev. Condens. Matter Phys.",
  publisher = "Annual Reviews",
  volume    =  7,
  number    =  1,
  pages     = "195--221",
  month     =  "10~" # mar,
  year      =  2016,
  url       = "https://doi.org/10.1146/annurev-conmatphys-031115-011319",
  doi       = "10.1146/annurev-conmatphys-031115-011319",
  issn      = "1947-5454"
}

@article{Moller2025-ly,
  title = {Rethinking $\alpha$-{RuCl}$_3$: Parameters, models, and phase diagram},
  author = {M\"oller, Marius and Maksimov, P. A. and Jiang, Shengtao and White, Steven R. and Valent\'{\i}, Roser and Chernyshev, A. L.},
  journal = {Phys. Rev. B},
  volume = {112},
  issue = {10},
  pages = {104403},
  numpages = {37},
  year = {2025},
  month = {Sep},
  publisher = {American Physical Society},
  doi = {10.1103/hflp-41lj},
  url = {https://link.aps.org/doi/10.1103/hflp-41lj}
}

@ARTICLE{Zhu2018-ve,
  title     = "Robust non-Abelian spin liquid and a possible intermediate phase
               in the antiferromagnetic Kitaev model with magnetic field",
  author    = "Zhu, Zheng and Kimchi, Itamar and Sheng, D N and Fu, Liang",
  journal   = "Phys. Rev. B.",
  publisher = "American Physical Society (APS)",
  volume    =  97,
  number    =  24,
  pages     =  241110,
  month     =  "15~" # jun,
  year      =  2018,
  url       = "https://link.aps.org/doi/10.1103/PhysRevB.97.241110",
  doi       = "10.1103/physrevb.97.241110",
  issn      = "2469-9950,2469-9969"
}

@ARTICLE{Hickey2019-yn,
  title   = "Emergence of a field-driven {U}(1) spin liquid in the Kitaev
             honeycomb model",
  author  = "Hickey, Ciarán and Trebst, Simon",
  journal = "Nat. Commun.",
  volume  =  10,
  number  =  1,
  pages   =  530,
  month   =  "31~" # jan,
  year    =  2019,
  url     = "http://dx.doi.org/10.1038/s41467-019-08459-9",
  doi     = "10.1038/s41467-019-08459-9",
  pmc     = "PMC6355955",
  pmid    =  30705263,
  issn    = "2041-1723"
}

@ARTICLE{Gohlke2018-kg,
  title     = "Dynamical and topological properties of the Kitaev model in a
               [111] magnetic field",
  author    = "Gohlke, Matthias and Moessner, Roderich and Pollmann, Frank",
  journal   = "Phys. Rev. B",
  publisher = "American Physical Society",
  volume    =  98,
  number    =  1,
  pages     =  014418,
  month     =  "17~" # jul,
  year      =  2018,
  url       = "https://link.aps.org/doi/10.1103/PhysRevB.98.014418",
  doi       = "10.1103/PhysRevB.98.014418",
  issn      = "0163-1829"
}

@ARTICLE{Gordon2019-iu,
  title   = "Theory of the field-revealed Kitaev spin liquid",
  author  = "Gordon, Jacob S and Catuneanu, Andrei and Sørensen, Erik S and Kee,
             Hae-Young",
  journal = "Nat. Commun.",
  volume  =  10,
  number  =  1,
  pages   =  2470,
  month   =  "6~" # jun,
  year    =  2019,
  url     = "http://dx.doi.org/10.1038/s41467-019-10405-8",
  doi     = "10.1038/s41467-019-10405-8",
  pmc     = "PMC6554414",
  pmid    =  31171777,
  issn    = "2041-1723"
}

@ARTICLE{Sorensen2021-jg,
  title     = "Heart of Entanglement: Chiral, Nematic, and Incommensurate Phases
               in the Kitaev-Gamma Ladder in a Field",
  author    = "S{\o}rensen, Erik S and Catuneanu, Andrei and Gordon, Jacob S and
               Kee, Hae-Young",
  journal   = "Phys. Rev. X",
  publisher = "American Physical Society",
  volume    =  11,
  number    =  1,
  pages     =  011013,
  month     =  "21~" # jan,
  year      =  2021,
  url       = "https://link.aps.org/doi/10.1103/PhysRevX.11.011013",
  doi       = "10.1103/PhysRevX.11.011013"
}

@article{Sorensen_2024_th,
  title = {Twice Hidden String Order and Competing Phases in the Spin-1/2 {{Kitaev}}--{{Gamma}} Ladder},
  author = {S{\o}rensen, Erik S. and Kee, Hae-Young},
  year = 2024,
  month = jan,
  journal = {npj Quantum Mater.},
  volume = {9},
  number = {1},
  pages = {10},
  publisher = {Nature Publishing Group},
  issn = {2397-4648},
  doi = {10.1038/s41535-024-00621-x},
  url = "https://www.nature.com/articles/s41535-024-00621-x"
}

@ARTICLE{Chern2020-ti,
  title     = "Magnetic field induced competing phases in spin-orbital entangled
               Kitaev magnets",
  author    = "Chern, Li Ern and Kaneko, Ryui and Lee, Hyun-Yong and Kim, Yong
               Baek",
  journal   = "Phys. Rev. Research",
  publisher = "American Physical Society",
  volume    =  2,
  number    =  1,
  pages     =  013014,
  month     =  "6~" # jan,
  year      =  2020,
  url       = "https://link.aps.org/doi/10.1103/PhysRevResearch.2.013014",
  doi       = "10.1103/PhysRevResearch.2.013014"
}

@ARTICLE{Lee2020-zx,
  title   = "Magnetic field induced quantum phases in a tensor network study of
             Kitaev magnets",
  author  = "Lee, Hyun-Yong and Kaneko, Ryui and Chern, Li Ern and Okubo,
             Tsuyoshi and Yamaji, Youhei and Kawashima, Naoki and Kim, Yong Baek",
  journal = "Nat. Commun.",
  volume  =  11,
  number  =  1,
  pages   =  1639,
  month   =  "2~" # apr,
  year    =  2020,
  url     = "http://dx.doi.org/10.1038/s41467-020-15320-x",
  doi     = "10.1038/s41467-020-15320-x",
  pmc     = "PMC7118087",
  pmid    =  32242020,
  issn    = "2041-1723"
}

@ARTICLE{Gohlke2020-go,
  title     = "Emergence of nematic paramagnet via quantum order-by-disorder and
               pseudo-Goldstone modes in Kitaev magnets",
  author    = "Gohlke, Matthias and Chern, Li Ern and Kee, Hae-Young and Kim,
               Yong Baek",
  journal   = "Phys. Rev. Res.",
  publisher = "American Physical Society (APS)",
  volume    =  2,
  number    =  4,
  pages     =  043023,
  month     =  "5~" # oct,
  year      =  2020,
  url       = "http://link.aps.org/pdf/10.1103/PhysRevResearch.2.043023",
  doi       = "10.1103/physrevresearch.2.043023",
  issn      = "2643-1564"
}

@article{Wang2024-yg,
  author  = {Wang, Jiucai and Normand, B. and Liu, Zheng-Xin},
  title   = {Multinode quantum spin liquids in extended {Kitaev} honeycomb models},
  journal = {npj Quantum Mater.},
  year    = {2024},
  volume  = {9},
  number  = {1},
  pages   = {94},
  doi     = {10.1038/s41535-024-00704-9},
  url     = {https://doi.org/10.1038/s41535-024-00704-9}
}

@ARTICLE{Wang2019-bz,
  title   = "One Proximate Kitaev Spin Liquid in the \protect\ensuremath{K}-\protect\ensuremath{J}-\protect\ensuremath{\Gamma} Model on the
             Honeycomb Lattice",
  author  = "Wang, Jiucai and Normand, B and Liu, Zheng-Xin",
  journal = "Phys. Rev. Lett.",
  volume  =  123,
  number  =  19,
  pages   =  197201,
  month   =  "8~" # nov,
  year    =  2019,
  url     = "http://dx.doi.org/10.1103/PhysRevLett.123.197201",
  doi     = "10.1103/PhysRevLett.123.197201",
  pmid    =  31765175,
  issn    = "0031-9007,1079-7114"
}

@ARTICLE{Chaloupka2015-uh,
  title     = "Hidden symmetries of the extended Kitaev-Heisenberg model:
               Implications for the honeycomb-lattice iridates
               {A}$_{2}${IrO}$_{3}$",
  author    = "Chaloupka, Jiří and Khaliullin, Giniyat",
  journal   = "Phys. Rev. B",
  publisher = "American Physical Society",
  volume    =  92,
  number    =  2,
  pages     =  024413,
  month     =  "13~" # jul,
  year      =  2015,
  url       = "https://link.aps.org/doi/10.1103/PhysRevB.92.024413",
  doi       = "10.1103/PhysRevB.92.024413",
  issn      = "0163-1829"
}

@ARTICLE{Rousochatzakis2024-td,
  title     = "Beyond Kitaev physics in strong spin-orbit coupled magnets",
  author    = "Rousochatzakis, Ioannis and Perkins, Natalia B and Luo, Qiang and
               Kee, Hae-Young",
  journal   = "Rep. Prog. Phys.",
  publisher = "IOP Publishing",
  volume    =  87,
  number    =  2,
  pages     =  026502,
  month     =  "2~" # feb,
  year      =  2024,
  url       = "https://iopscience.iop.org/article/10.1088/1361-6633/ad208d/meta",
  doi       = "10.1088/1361-6633/ad208d",
  pmid      =  38241723,
  issn      = "0034-4885,1361-6633"
}

@ARTICLE{Yokoi2021-ul,
  title   = "Half-integer quantized anomalous thermal Hall effect in the Kitaev
             material candidate $\alpha$-{RuCl}$_{3}$",
  author  = "Yokoi, T and Ma, S and Kasahara, Y and Kasahara, S and Shibauchi, T
             and Kurita, N and Tanaka, H and Nasu, J and Motome, Y and Hickey, C
             and Trebst, S and Matsuda, Y",
  journal = "Science",
  volume  =  373,
  number  =  6554,
  pages   = "568--572",
  month   =  "30~" # jul,
  year    =  2021,
  url     = "http://dx.doi.org/10.1126/science.aay5551",
  doi     = "10.1126/science.aay5551",
  pmid    =  34326240,
  issn    = "0036-8075,1095-9203"
}

@ARTICLE{Kasahara2018-or,
  title     = "Majorana quantization and half-integer thermal quantum Hall
               effect in a Kitaev spin liquid",
  author    = "Kasahara, Y and Ohnishi, T and Mizukami, Y and Tanaka, O and Ma,
               Sixiao and Sugii, K and Kurita, N and Tanaka, H and Nasu, J and
               Motome, Y and Shibauchi, T and Matsuda, Y",
  journal   = "Nature",
  publisher = "Springer US",
  volume    =  559,
  number    =  7713,
  pages     = "227--231",
  month     =  jul,
  year      =  2018,
  url       = "http://dx.doi.org/10.1038/s41586-018-0274-0",
  eprint    = "1805.05022",
  doi       = "10.1038/s41586-018-0274-0",
  pmid      =  29995863,
  issn      = "0028-0836,1476-4687"
}

@ARTICLE{Czajka2023-da,
  title     = "Planar thermal Hall effect of topological bosons in the Kitaev
               magnet $\alpha$-{RuCl}$_3$",
  author    = "Czajka, Peter and Gao, Tong and Hirschberger, Max and
               Lampen-Kelley, Paula and Banerjee, Arnab and Quirk, Nicholas and
               Mandrus, David G and Nagler, Stephen E and Ong, N P",
  journal   = "Nat. Mater.",
  publisher = "Springer Science and Business Media LLC",
  volume    =  22,
  number    =  1,
  pages     = "36--41",
  month     =  jan,
  year      =  2023,
  url       = "http://dx.doi.org/10.1038/s41563-022-01397-w",
  doi       = "10.1038/s41563-022-01397-w",
  pmid      =  36396962,
  issn      = "1476-1122,1476-4660"
}

@ARTICLE{Czajka2021-tm,
  title     = "Oscillations of the thermal conductivity in the spin-liquid state
               of $\alpha$-{RuCl}$_3$",
  author    = "Czajka, Peter and Gao, Tong and Hirschberger, Max and
               Lampen-Kelley, Paula and Banerjee, Arnab and Yan, Jiaqiang and
               Mandrus, David G and Nagler, Stephen E and Ong, N P",
  journal   = "Nat. Phys.",
  publisher = "Nature Publishing Group",
  volume    =  17,
  number    =  8,
  pages     = "915--919",
  month     =  "13~" # may,
  year      =  2021,
  url       = "https://www.nature.com/articles/s41567-021-01243-x",
  doi       = "10.1038/s41567-021-01243-x",
  issn      = "1745-2473"
}

@ARTICLE{Bruin2022-wq,
  title     = "Robustness of the thermal Hall effect close to half-quantization
               in $\alpha$-{RuCl}$_3$",
  author    = "Bruin, J A N and Claus, R R and Matsumoto, Y and Kurita, N and
               Tanaka, H and Takagi, H",
  journal   = "Nat. Phys.",
  publisher = "Nature Publishing Group",
  volume    =  18,
  number    =  4,
  pages     = "401--405",
  month     =  "17~" # feb,
  year      =  2022,
  url       = "https://www.nature.com/articles/s41567-021-01501-y",
  doi       = "10.1038/s41567-021-01501-y",
  issn      = "1745-2473"
}

@ARTICLE{Maksimov2020-up,
  title     = "Rethinking $\alpha$-{RuCl}$_{3}$",
  author    = "Maksimov, P A and Chernyshev, A L",
  journal   = "Phys. Rev. Research",
  publisher = "American Physical Society",
  volume    =  2,
  number    =  3,
  pages     =  033011,
  month     =  "2~" # jul,
  year      =  2020,
  url       = "https://link.aps.org/doi/10.1103/PhysRevResearch.2.033011",
  doi       = "10.1103/PhysRevResearch.2.033011"
}

@article{Jiefu2025-wi,
  title = {Intermediate phases in $\alpha$-{RuCl}$_{3}$ under in-plane magnetic field via interlayer spin interactions},
  author = {Cen, Jiefu and Kee, Hae-Young},
  journal = {Phys. Rev. B},
  volume = {112},
  issue = {2},
  pages = {024419},
  numpages = {12},
  year = {2025},
  month = {Jul},
  publisher = {American Physical Society},
  doi = {10.1103/tr8c-pxbr},
  url = {https://link.aps.org/doi/10.1103/tr8c-pxbr}
}

@ARTICLE{Li2021-gy,
  title   = "Identification of magnetic interactions and high-field quantum spin
             liquid in $\alpha$-{RuCl$_3$}",
  author  = "Li, Han and Zhang, Hao-Kai and Wang, Jiucai and Wu, Han-Qing and
             Gao, Yuan and Qu, Dai-Wei and Liu, Zheng-Xin and Gong, Shou-Shu and
             Li, Wei",
  journal = "Nat. Commun.",
  volume  =  12,
  number  =  1,
  pages   =  4007,
  month   =  "29~" # jun,
  year    =  2021,
  url     = "http://dx.doi.org/10.1038/s41467-021-24257-8",
  doi     = "10.1038/s41467-021-24257-8",
  pmc     = "PMC8242101",
  pmid    =  34188044,
  issn    = "2041-1723"
}

@ARTICLE{Chen2025-be,
  title     = "Anyon polarons as a window into competing phases of the Kitaev
               honeycomb model under a Zeeman field",
  author    = "Chen, Chuan and Villadiego, Inti Sodemann",
  journal   = "Phys. Rev. B",
  publisher = "American Physical Society (APS)",
  volume    =  111,
  number    =  24,
  pages     =  245140,
  month     =  "20~" # jun,
  year      =  2025,
  url       = "http://dx.doi.org/10.1103/xxym-pc1t",
  doi       = "10.1103/xxym-pc1t",
  issn      = "2469-9950,2469-9969"
}

@ARTICLE{Chen2023-vo,
  title     = "Nature of visons in the perturbed ferromagnetic and
               antiferromagnetic Kitaev honeycomb models",
  author    = "Chen, Chuan and Villadiego, Inti Sodemann",
  journal   = "Phys. Rev. B",
  publisher = "American Physical Society (APS)",
  volume    =  107,
  number    =  4,
  pages     =  045114,
  month     =  "10~" # jan,
  year      =  2023,
  url       = "https://link.aps.org/doi/10.1103/PhysRevB.107.045114",
  doi       = "10.1103/physrevb.107.045114",
  issn      = "2469-9950,2469-9969"
}

@article{Joy2022-ba,
  title = {Dynamics of Visons and Thermal Hall Effect in Perturbed Kitaev Models},
  author = {Joy, Aprem P. and Rosch, Achim},
  journal = {Phys. Rev. X},
  volume = {12},
  issue = {4},
  pages = {041004},
  numpages = {19},
  year = {2022},
  month = {Oct},
  publisher = {American Physical Society},
  doi = {10.1103/PhysRevX.12.041004},
  url = {https://link.aps.org/doi/10.1103/PhysRevX.12.041004}
}

@ARTICLE{Zhang2022-tc,
  title   = "Theory of the Kitaev model in a [111] magnetic field",
  author  = "Zhang, Shang-Shun and Halász, Gábor B and Batista, Cristian D",
  journal = "Nat. Commun.",
  volume  =  13,
  number  =  1,
  pages   =  399,
  month   =  "20~" # jan,
  year    =  2022,
  url     = "http://dx.doi.org/10.1038/s41467-022-28014-3",
  doi     = "10.1038/s41467-022-28014-3",
  pmc     = "PMC8776843",
  pmid    =  35058433,
  issn    = "2041-1723"
}

@article{Zhang2021-xi,
  title = {Variational study of the Kitaev-Heisenberg-Gamma model},
  author = {Zhang, Shang-Shun and Hal\'asz, G\'abor B. and Zhu, Wei and Batista, Cristian D.},
  journal = {Phys. Rev. B},
  volume = {104},
  issue = {1},
  pages = {014411},
  numpages = {18},
  year = {2021},
  month = {Jul},
  publisher = {American Physical Society},
  doi = {10.1103/PhysRevB.104.014411},
  url = {https://link.aps.org/doi/10.1103/PhysRevB.104.014411}
}

@ARTICLE{Knolle2014-wl,
  title     = "Dynamics of a Two-Dimensional Quantum Spin Liquid: Signatures of
               Emergent Majorana Fermions and Fluxes",
  author    = "Knolle, J and Kovrizhin, D L and Chalker, J T and Moessner, R",
  journal   = "Phys. Rev. Lett.",
  publisher = "American Physical Society",
  volume    =  112,
  number    =  20,
  pages     =  207203,
  month     =  "21~" # may,
  year      =  2014,
  url       = "https://link.aps.org/doi/10.1103/PhysRevLett.112.207203",
  doi       = "10.1103/PhysRevLett.112.207203",
  issn      = "0031-9007"
}

@ARTICLE{Knolle2015-lx,
  title     = "Dynamics of fractionalization in quantum spin liquids",
  author    = "Knolle, J and Kovrizhin, D L and Chalker, J T and Moessner, R",
  journal   = "Phys. Rev. B",
  publisher = "American Physical Society",
  volume    =  92,
  number    =  11,
  pages     =  115127,
  month     =  "14~" # sep,
  year      =  2015,
  url       = "https://link.aps.org/doi/10.1103/PhysRevB.92.115127",
  doi       = "10.1103/PhysRevB.92.115127",
  issn      = "0163-1829"
}

@ARTICLE{Sears2020-si,
  title     = "Ferromagnetic Kitaev interaction and the origin of large magnetic
               anisotropy in $\alpha$-{RuCl}$_3$",
  author    = "Sears, Jennifer A and Chern, Li Ern and Kim, Subin and
               Bereciartua, Pablo J and Francoual, Sonia and Kim, Yong Baek and
               Kim, Young-June",
  journal   = "Nat. Phys.",
  publisher = "Nature Publishing Group",
  volume    =  16,
  number    =  8,
  pages     = "837--840",
  month     =  "13~" # apr,
  year      =  2020,
  url       = "https://www.nature.com/articles/s41567-020-0874-0",
  doi       = "10.1038/s41567-020-0874-0",
  issn      = "1745-2473"
}

@article{Rao2025-pp,
  title = {Dynamical response theory of interacting Majorana fermions and its application to generic Kitaev quantum spin liquids in a field},
  author = {Rao, Peng and Moessner, Roderich and Knolle, Johannes},
  journal = {Phys. Rev. B},
  volume = {112},
  issue = {2},
  pages = {024440},
  numpages = {18},
  year = {2025},
  month = {Jul},
  publisher = {American Physical Society},
  doi = {10.1103/f4b4-h1yr},
  url = {https://link.aps.org/doi/10.1103/f4b4-h1yr}
}

@article{Joy2024-cx,
  title = {Gauge Field Dynamics in Multilayer {{Kitaev}} Spin Liquids},
  author = {Joy, Aprem P. and Rosch, Achim},
  year = 2024,
  month = aug,
  journal = {npj Quantum Mater.},
  volume = {9},
  number = {1},
  pages = {62},
  issn = {2397-4648},
  doi = {10.1038/s41535-024-00673-z},
  url = "https://www.nature.com/articles/s41535-024-00673-z"
}

@article{Joy2025-pk,
  title = {Raman spectroscopy of anyons in generic Kitaev spin liquids},
  author = {Joy, Aprem P. and Rosch, Achim},
  journal = {Phys. Rev. B},
  volume = {112},
  issue = {18},
  pages = {184411},
  numpages = {16},
  year = {2025},
  month = {Nov},
  publisher = {American Physical Society},
  doi = {10.1103/b175-chx4},
  url = {https://link.aps.org/doi/10.1103/b175-chx4}
}

@ARTICLE{Zschocke2015-zr,
  title     = "Physical states and finite-size effects in Kitaev's honeycomb
               model: Bond disorder, spin excitations, and {NMR} line shape",
  author    = "Zschocke, Fabian and Vojta, Matthias",
  journal   = "Phys. Rev. B",
  publisher = "American Physical Society",
  volume    =  92,
  number    =  1,
  pages     =  014403,
  month     =  "6~" # jul,
  year      =  2015,
  url       = "http://link.aps.org/pdf/10.1103/PhysRevB.92.014403",
  doi       = "10.1103/PhysRevB.92.014403"
}

@ARTICLE{Hermele2005-ge,
  title     = "Algebraic spin liquid as the mother of many competing orders",
  author    = "Hermele, Michael and Senthil, T and Fisher, Matthew P A",
  journal   = "Phys. Rev. B",
  publisher = "American Physical Society",
  volume    =  72,
  number    =  10,
  pages     =  104404,
  month     =  "2~" # sep,
  year      =  2005,
  url       = "https://link.aps.org/doi/10.1103/PhysRevB.72.104404",
  doi       = "10.1103/PhysRevB.72.104404",
  issn      = "0163-1829"
}

@ARTICLE{Chen2022-tr,
  title     = "Berry phases of vison transport in {Z}$_{2}$ topologically
               ordered states from exact fermion-flux lattice dualities",
  author    = "Chen, Chuan and Rao, Peng and Sodemann, Inti",
  journal   = "Phys. Rev. Research",
  publisher = "American Physical Society",
  volume    =  4,
  number    =  4,
  pages     =  043003,
  month     =  "4~" # oct,
  year      =  2022,
  url       = "https://link.aps.org/doi/10.1103/PhysRevResearch.4.043003",
  doi       = "10.1103/PhysRevResearch.4.043003"
}

@ARTICLE{Zou2020-ni,
  title     = "Field-induced {QCD}$_{3}$-Chern-Simons quantum criticalities in
               Kitaev materials",
  author    = "Zou, Liujun and He, Yin-Chen",
  journal   = "Phys. Rev. Res.",
  publisher = "American Physical Society",
  volume    =  2,
  number    =  1,
  pages     =  013072,
  month     =  "23~" # jan,
  year      =  2020,
  url       = "https://link.aps.org/doi/10.1103/PhysRevResearch.2.013072",
  doi       = "10.1103/PhysRevResearch.2.013072"
}

@ARTICLE{Lee2006-mw,
  title     = "Doping a Mott insulator: Physics of high-temperature
               superconductivity",
  author    = "Lee, Patrick A and Nagaosa, Naoto and Wen, Xiao Gang",
  journal   = "Rev. Mod. Phys.",
  publisher = "American Physical Society",
  volume    =  78,
  number    =  1,
  pages     = "17--85",
  month     =  "6~" # jan,
  year      =  2006,
  url       = "https://link.aps.org/doi/10.1103/RevModPhys.78.17",
  eprint    = "cond-mat/0410445",
  doi       = "10.1103/RevModPhys.78.17",
  issn      = "0034-6861"
}

@ARTICLE{Burnell2018-fm,
  title     = "Anyon Condensation and Its Applications",
  author    = "Burnell, F J",
  journal   = "Annu. Rev. Condens. Matter Phys.",
  publisher = "Annual Reviews",
  volume    =  9,
  number    =  1,
  pages     = "307--327",
  month     =  "10~" # mar,
  year      =  2018,
  url       = "https://doi.org/10.1146/annurev-conmatphys-033117-054154",
  doi       = "10.1146/annurev-conmatphys-033117-054154",
  issn      = "1947-5454"
}

@ARTICLE{Pozo2021-gy,
  title     = "Anatomy of {Z}$_{2}$ fluxes in anyon Fermi liquids and Bose
               condensates",
  author    = "Pozo, Oscar and Rao, Peng and Chen, Chuan and Sodemann, Inti",
  journal   = "Phys. Rev. B",
  publisher = "American Physical Society",
  volume    =  103,
  number    =  3,
  pages     =  035145,
  month     =  "28~" # jan,
  year      =  2021,
  url       = "https://link.aps.org/doi/10.1103/PhysRevB.103.035145",
  doi       = "10.1103/PhysRevB.103.035145",
  issn      = "0163-1829"
}

@ARTICLE{Kitaev2006-ky,
  title     = "Topological entanglement entropy",
  author    = "Kitaev, Alexei and Preskill, John",
  journal   = "Phys. Rev. Lett.",
  publisher = "American Physical Society (APS)",
  volume    =  96,
  number    =  11,
  pages     =  110404,
  month     =  "24~" # mar,
  year      =  2006,
  url       = "http://dx.doi.org/10.1103/PhysRevLett.96.110404",
  doi       = "10.1103/PhysRevLett.96.110404",
  pmid      =  16605802,
  issn      = "0031-9007,1079-7114"
}

@ARTICLE{Levin2006-zx,
  title     = "Detecting topological order in a ground state wave function",
  author    = "Levin, Michael and Wen, Xiao-Gang",
  journal   = "Phys. Rev. Lett.",
  publisher = "American Physical Society (APS)",
  volume    =  96,
  number    =  11,
  pages     =  110405,
  month     =  "24~" # mar,
  year      =  2006,
  url       = "http://dx.doi.org/10.1103/PhysRevLett.96.110405",
  doi       = "10.1103/PhysRevLett.96.110405",
  pmid      =  16605803,
  issn      = "0031-9007,1079-7114"
}

@ARTICLE{Chen2018-nq,
  title   = "Exact bosonization in two spatial dimensions and a new class of
             lattice gauge theories",
  author = "Chen, Yu-An and Kapustin, Anton and Radi{\v{c}}evi{\'c}, {\DJ}or{\dj}e",
  journal = "Ann. Phys.",
  volume  =  393,
  pages   = "234--253",
  month   =  jun,
  year    =  2018,
  url     = "https://linkinghub.elsevier.com/retrieve/pii/S0003491618300873",
  eprint  = "1711.00515",
  doi     = "10.1016/j.aop.2018.03.024",
  issn    = "0003-4916"
}

\clearpage


\section*{Supplementary material}

# Anyon polarons as a window into the competing phases of the Kitaev-Gamma-Gamma' model

Chuan Chen[1, 2, *] and Inti Sodemann Villadiego[3, †]

[1]*School of Physical Science and Technology, Lanzhou University, Lanzhou 730000, China*
[2]*Lanzhou Center for Theoretical Physics, Key Laboratory of Quantum Theory and Applications of MoE,*
*Key Laboratory of Theoretical Physics of Gansu Province,*
*Gansu Provincial Research Center for Basic Disciplines of Quantum Physics, Lanzhou University, Lanzhou 730000, China*
[3]*Institut für Theoretische Physik, Universität Leipzig, 04103 Leipzig, Germany*

## CONTENTS

## S-I. SINGLE VISON HOPPING

In this section, we discuss the $\Gamma'$-induced linear-order single vison hopping, and the higher-order effects of the $\Gamma$ interaction on single visons.

### S-I.A. $\Gamma'$-induced single-vison hopping

Here we explain the $\Gamma'$-induced single-vison hopping through two examples illustrated in Fig. S1. Note that there is another fixed vison to the left end (not shown) in both two cases, and the red lines connecting the two visons indicate the branch cut (BC), where the each bond crossed by a red line is occupied by a $\chi$ fermion.

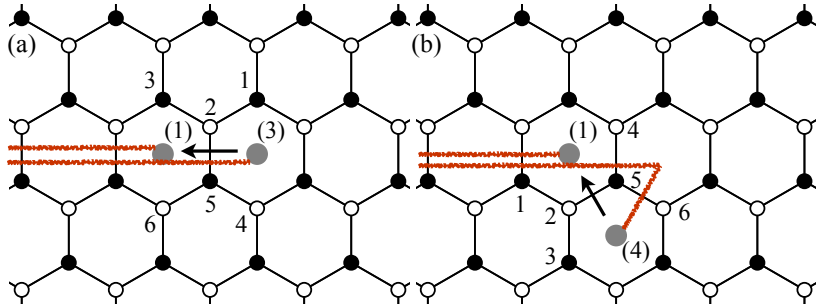

FIG. S1. Two different single-vison hopping processes induced by the $\Gamma'$ interaction. (a) Hopping of a vison from $(3) \to (1)$. (b) Hopping of a vison from $(4) \to (1)$. Note in both cases, there is another vison located at the far left end of the system, which is not shown explicitly. The red linds indicates the BC associated with the visons. The bonds crossed by the red line are occupied by the gauge $\chi$ fermions.

---

* chenc@lzu.edu.cn
† sodemann@itp.uni-leipzig.de

I). The hopping of a vison from location (3) to location (1), as shown in Fig. S1(a). For state associated with configuration (1), one can choose it to contain an *even* number of $\chi$ fermions:

$$|\Psi(1)\rangle = \mathcal{N} P \left[ \prod_{(\boldsymbol{r},\alpha) \in \mathrm{BC}(1)} \chi^\dagger_{\boldsymbol{r},\alpha} \right] |0_\chi; \Psi^{\mathrm{even}}_c(1)\rangle. \tag{S1}$$

Here $P = \prod_r (1 + D_r)/2$ with $D_r = b^x_r b^y_r b^z_r c_r$ is the projection operator. The normalization factor $\mathcal{N} = 2^{N-1/2}$ where $N$ is the number of unit cells in the (periodic) system. To guarantee that $|\Psi(1)\rangle$ is physical, $|\Psi^0_c(1)\rangle$ should also contain an even number of matter $a$ fermions, with $a_{\boldsymbol{r}} = c_{\boldsymbol{r}} + ic_{\boldsymbol{r}+\boldsymbol{\delta}_x}$, $\boldsymbol{r} \in A$ [1]. Two scenarios may arise: (i). The effective $c$-fermion BdG Hamiltonian $H_c[(1)]$, corresponding to the $\mathbb{Z}_2$ gauge-field configuration (1), has a ground state with even $a$-fermion parity. In this case, the ground state can be directly taken as $|\Psi^{\mathrm{even}}_c(1)\rangle$; (ii). The ground state of $H_c[(1)]$ has odd $a$ parity, thus unphysical. Then, the first excited state with even parity should be used as $|\Psi^{\mathrm{even}}_c(1)\rangle$ instead. For case (ii), one can define a new BdG Hamiltonian $H'_c[(1)]$, such that the even-parity parity $|\Psi^{\mathrm{even}}_c(1)\rangle$ is its ground state and can be represented in a BCS form.

For state associated with configuration (3), since there is an additional $\chi_{5,z}$ bond fermion compared to $|\Psi(1)\rangle$, it must contains an odd number of $a$ fermions. It can be written as:

$$|\Psi(3)\rangle = \mathcal{N} P \left[ \prod_{(\boldsymbol{r},\alpha) \in \mathrm{BC}(1)} \chi^\dagger_{\boldsymbol{r},\alpha} \right] \chi^\dagger_{5,z} \alpha^\dagger_1(3) |0_\chi; \Psi^{\mathrm{even}}_c(3)\rangle. \tag{S2}$$

The idea is to write the matter fermion part as the first excited state of a BdG Hamiltonian whose ground state has even $a$-parity. There are two possible scenarios: (i). The ground state of $H_c[(3)]$ has even $a$-parity. Then one simply take $|\Psi^{\mathrm{even}}_c(3)\rangle$ to be the ground state of $H_c[(3)]$, and $\alpha^\dagger_1(3)$ to be the creation operator for the first (lowest-energy) Bogoliubov quasiparticle (Bogolon) of it. (ii). The ground state of $H_c[(3)]$ has odd $a$-parity. In this case, the matter fermion part is the ground state of $H_c[(3)]$. To represent $|\Psi^{\mathrm{even}}_c(3)\rangle$ in a BCS form, one can define a new BdG Hamiltonian $H'_c[(3)]$, so that the first excited state of $H_c[(3)]$, which is $|\Psi^{\mathrm{even}}_c(3)\rangle$, is the *ground state* of $H'_c[(3)]$. $\alpha^\dagger_1(3)$ creates the first Bogolon of $H'_c[(3)]$.

There are 4 terms from the $\Gamma'$ interaction that contribute to this hopping process:

$$\sigma^y_1 \sigma^z_2 \leftrightarrow (b^x_5 c_5)(ic_5 c_2)(-ib^x_5 b^x_2)(ic_1 c_2)(-ib^y_1 b^y_2), \tag{S3a}$$

$$\sigma^z_3 \sigma^y_2 \leftrightarrow (-b^x_5 c_5)(ic_5 c_2)(-ib^x_5 b^x_2)(ic_3 c_2)(-ib^z_3 b^z_2), \tag{S3b}$$

$$\sigma^y_5 \sigma^z_4 \leftrightarrow (-b^x_5 c_5)(ic_5 c_4)(-ib^z_5 b^z_4), \tag{S3c}$$

$$\sigma^z_5 \sigma^y_6 \leftrightarrow (b^x_5 c_5)(ic_5 c_6)(-ib^y_5 b^y_6). \tag{S3d}$$

After some algebra, it can be shown that the $\Gamma'$-induced hopping for this process reads:

$$\Gamma' \langle 0_\chi; \Psi^{\mathrm{even}}_c(1) | (c_1 + c_3 - ic_4 + ic_6) \alpha^\dagger_1(3) | 0_\chi; \Psi^{\mathrm{even}}_c(3) \rangle, \tag{S4}$$

which can be calculated using standard method [2–6].

II). The hopping from (4) to (1), as shown in Fig. S1(b). Since the state $|\Psi(4)\rangle$ has two more gauge fermions ($\chi_{5,z}$ and $\chi_{5,y}$) than $|\Psi(1)\rangle$, its fermion part should also has even $a$-parity. It can be written as:

$$|\Psi(4)\rangle = \mathcal{N} P \left[ \prod_{(\boldsymbol{r},\alpha) \in \mathrm{BC}(1)} \chi^\dagger_{\boldsymbol{r},\alpha} \right] \chi^\dagger_{5,z} \chi^\dagger_{5,y} |0_\chi; \Psi^{\mathrm{even}}_c(4)\rangle. \tag{S5}$$

There are two possible cases: (i). The BdG Hamiltonian $H_c[(4)]$'s ground state has even parity. In this case $|\Psi^{\mathrm{even}}_c(4)\rangle$ can be taken to be the ground state of it. (ii). $H_c[(4)]$'s ground state has odd parity. In this case, $|\Psi^{\mathrm{even}}_c(4)\rangle$ is the first excited state of $H_c[(4)]$. One can define a new Hamiltonian $H'_c[(4)]$, such that $|\Psi^{\mathrm{even}}_c(4)\rangle$ is its ground state and can be written in a BCS form.

There are 4 terms contribute to this hopping process:

$$\sigma^y_5 \sigma^z_4 \leftrightarrow (b^y_5 b^z_5)(ic_5 c_4)(-ib^z_5 b^z_4), \tag{S6a}$$

$$\sigma^z_5 \sigma^y_6 \leftrightarrow (-b^y_5 b^z_5)(ic_5 c_6)(-ib^y_5 b^y_6), \tag{S6b}$$

$$\sigma^z_2 \sigma^y_1 \leftrightarrow (-b^y_5 b^z_5)(ic_5 c_2)(-ib^x_5 b^x_2)(ic_1 c_2)(-ib^y_1 b^y_2), \tag{S6c}$$

$$\sigma^y_2 \sigma^z_3 \leftrightarrow (b^y_5 b^z_5)(ic_5 c_2)(-ib^x_5 b^x_2)(ic_3 c_2)(-ib^z_3 b^z_2). \tag{S6d}$$

After some algebra, it can be shown that the hopping amplitude for this process reads:

$$\Gamma' \langle 0_\chi; \Psi^{\mathrm{even}}_c(1) | c_5 (-c_1 + c_3 - ic_4 + ic_6) | 0_\chi; \Psi^{\mathrm{even}}_c(4) \rangle. \tag{S7}$$

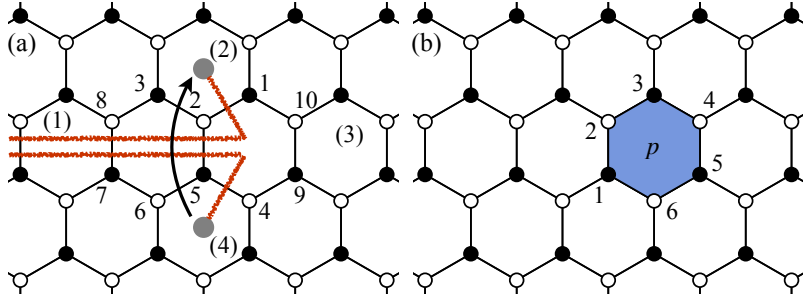

FIG. S2. (a) Γ-induced second-order single-vison hopping. Note that there is another vison fixed to the left end, and the red lines connecting the fixed and mobile visons indicate the BC. For any bond crossed by the red line, there is an occupied bond $\chi$ fermion. (b) Γ-induced third-order single-vison chemical potential term.

### S-I.B.  Higher-order effects of the Γ interaction

In this section, we discuss the higher-order effects of the Γ interaction on single visons.

First, we discuss the second-order single-vison hopping. We take the example shown in Fig. S2(a) where a single vison hops from (4) → (2). To perform a perturbative analysis, the Hilbert space can be divided into a low-energy subspace containing two visons (one fixed and one mobile), and a high-energy subspace involving additional vison excitations. The at second order, the low-energy effective Hamiltonian reads:

$$P_L H_\Gamma P_H \frac{1}{E_0 - H_K} P_H H_\Gamma P_L \approx \frac{1}{-\Delta E} P_L H_\Gamma P_H H_\Gamma P_L. \tag{S8}$$

Here $P_L$ ($P_H$) is the projection operator into the low-energy (high-energy) subspace. We define $\Delta E = E_H - E_0$, where $E_H$ is the energy of the *lowest* high-energy state. This definition may lead to an overestimation of the prefactor of the interaction. One can show that the intermediate state resulting from a single Γ term contains three visons, located either at plaquettes (4), (3), (2) or at (4), (1), (2). It can be shown that the second-order interaction that induce the hopping reads:

$$\frac{\Gamma^2}{-\Delta E} \left[ (\sigma_7^x \sigma_6^z + \sigma_7^z \sigma_6^x)(\sigma_3^z \sigma_8^y + \sigma_3^y \sigma_8^z) + (\sigma_9^y \sigma_4^z + \sigma_9^z \sigma_4^y)(\sigma_1^z \sigma_{10}^x + \sigma_1^x \sigma_{10}^z) \right]. \tag{S9}$$

Evaluating the matrix element of this term between the initial state $|\Psi(4)\rangle$ and the final state $|\Psi(2)\rangle$ reveals that it *vanishes* in the AFM Kitaev model but remains finite in the FM Kitaev model.

Now we discuss the vison chemical potential term generated at third-order [see Fig. S2(b)]. At third-order, the effective Hamiltonian has the form:

$$P_L H_\Gamma P_H \frac{1}{E_0 - H_K} H_\Gamma \frac{1}{E_0 - H_K} P_H H_\Gamma P_L \approx \frac{1}{\Delta E^2} P_L H_\Gamma P_H H_\Gamma P_H H_\Gamma P_L. \tag{S10}$$

Here $\Delta E \approx 0.23|K|$ is the energy difference between the (lowest) intermediate state and the initial state, which can lead to an overestimation of the overall factor. It can be shown that a vison chemical potential term is generated for each plaquette $p$:

$$\frac{\Gamma^3}{\Delta E^2} 12 W_p = \frac{\Gamma^3}{\Delta E^2} 12(1 - 2n_p^v). \tag{S11}$$

Here $W_p = \sigma_1^x \sigma_2^y \sigma_3^z \sigma_4^x \sigma_5^y \sigma_6^z$ is the vison parity for plaquette $p$, and $n_p^v$ is the vison number operator. This term favors (disfavors) the vison excitation for $\Gamma > 0$ ($\Gamma < 0$). Since the prefactor of the chemical potential term is likely overestimated by this approach and an accurate estimate is difficult to obtain accurately, we have used $-0.3 \frac{\Gamma^3}{\Delta E^2} n_p^v$ to plot the gray dashed line in Fig. 7(b) of the main text.

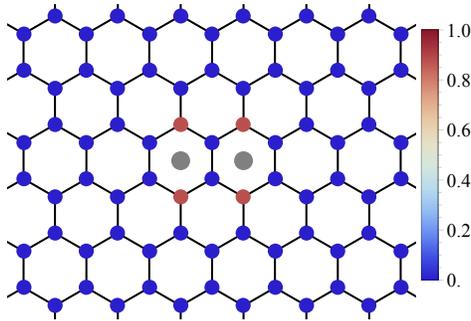

FIG. S3. Coupling between a $\tilde{\chi}_{\boldsymbol{r},z}$ vison pair (two visons are represented by the gray dots) and nearby $c$ fermions. It couples strongly to the four $c$ fermions nearby (shown in red color), with coupling strength around 0.88. The couplings to all the other $c$ fermions (shown in blue color) are vanishingly small. Notably, the result is identical for both FM and AFM Kitaev models.

## S-II.  COUPLING BETWEEN $\tilde{\chi}$ AND $c$ FERMIONS

In this section, we discuss the determination of the $\Gamma'$-induced coupling $(p_{\boldsymbol{R},\alpha})$ between $\tilde{\chi}$ and $c$ fermions in the effective fermion Hamiltonian $H_{\rm f}$. The coupling between $\tilde{\chi}$ and $c$ can be written as [6]:

$$\sum_{\boldsymbol{r},\alpha}\sum_{j} \lambda_j(\boldsymbol{r},\alpha)\alpha_j^\dagger \tilde{\chi}_{\boldsymbol{r},\alpha} + \eta_j(\boldsymbol{r},\alpha)\tilde{\chi}_{\boldsymbol{r},\alpha}\tilde{\alpha}_j(\boldsymbol{r},\alpha) + {\rm H.c.} \tag{S12}$$

Here $\alpha_j^\dagger$ creates the $j$-th Bogolon of the zero-flux $c$-fermion BdG Hamiltonian $H_c(0)$, and $\tilde{\alpha}_j(\boldsymbol{r},\alpha)^\dagger$ creates the $j$-th Bogolon of the BdG Hamiltonian $H_c(\boldsymbol{r},\alpha)$, which corresponds to the configuration with two visons adjacent to the bond $(\boldsymbol{r},\alpha)$. The coupling constants $\lambda_j(\boldsymbol{r},\alpha)$ and $\eta_j(\boldsymbol{r},\alpha)$ can be determined through:

$$\lambda_j(\boldsymbol{r},\alpha) = \langle\Omega|\alpha_j H_{\Gamma'}\tilde{\chi}_{\boldsymbol{r},\alpha}^\dagger|\Omega\rangle, \tag{S13a}$$

$$\eta_j(\boldsymbol{r},\alpha) = \langle\Omega|H_{\Gamma'}\tilde{\alpha}_j^\dagger(\boldsymbol{r},\alpha)\tilde{\chi}_{\boldsymbol{r},\alpha}^\dagger|\Omega\rangle. \tag{S13b}$$

After obtaining all the $\lambda_j$ and $\eta_j$, one can express $\alpha_j$ and $\tilde{\alpha}j(\boldsymbol{r},\alpha)$ in terms of the local $c$ fermions, thereby deriving the coupling between the $c$ fermions and the $\tilde{\chi}\boldsymbol{r},\alpha$ fermions, as presented in the main text. Fig. S3 shows the coupling between a $\tilde{\chi}\boldsymbol{r},z$ fermion and the surrounding $c$ fermions. The coupling is strongest with the four nearest $c$ fermions (highlighted in red) and vanishes for all others. Interestingly, the value of $p_{\boldsymbol{R},\alpha}/\Gamma'$ is independent of the sign of the Kitaev interaction. Moreover, due to the specific form of the real-space coupling, the two remain decoupled at the K point in momentum space, and the Dirac cone of the $c$ fermions is preserved.

[1] F. Zschocke and M. Vojta, Physical states and finite-size effects in kitaev's honeycomb model: Bond disorder, spin excitations, and NMR line shape, Phys. Rev. B **92**, 014403 (2015).
[2] S.-S. Zhang, G. B. Halász, W. Zhu, and C. D. Batista, Variational study of the kitaev-heisenberg-gamma model, Phys. Rev. B **104**, 014411 (2021).
[3] S.-S. Zhang, G. B. Halász, and C. D. Batista, Theory of the kitaev model in a [111] magnetic field, Nat. Commun. **13**, 399 (2022).
[4] A. P. Joy and A. Rosch, Dynamics of visons and thermal hall effect in perturbed kitaev models, Phys. Rev. X **12**, 041004 (2022).
[5] C. Chen and I. S. Villadiego, Nature of visons in the perturbed ferromagnetic and antiferromagnetic kitaev honeycomb models, Phys. Rev. B **107**, 045114 (2023).
[6] C. Chen and I. S. Villadiego, Anyon polarons as a window into competing phases of the kitaev honeycomb model under a zeeman field, Phys. Rev. B **111**, 245140 (2025).



\end{document}